\documentclass[twocolumn, twocolappendix,times]{aastex63}

\shorttitle{Quantifying Feedback from Narrow Line Region Outflows}
\shortauthors{Revalski et al.}

\usepackage{float}
\usepackage{color}
\usepackage{comment}
\usepackage{graphics}
\usepackage{epstopdf}
\usepackage{microtype}
\usepackage{subfigure}
\usepackage[normalem]{ulem}
\usepackage{natbib}
\graphicspath{{./}{figures/}}
\usepackage{amsmath}
\newcommand{\lbol}{\ensuremath{L_{bol}}}
\newcommand{\mbh}{\ensuremath{M_{BH}}}
\newcommand{\ledd}{\ensuremath{\lbol/L_{Edd}}}

\begin{document}

\title{Quantifying Feedback from Narrow Line Region Outflows in Nearby Active Galaxies. IV.\\The Effects of Different Density Estimates on the Ionized Gas Masses and Outflow Rates}

\correspondingauthor{Mitchell Revalski}
\email{mrevalski@stsci.edu}

\author[0000-0002-4917-7873]{Mitchell Revalski}
\affiliation{Space Telescope Science Institute, 3700 San Martin Drive, Baltimore, MD 21218, USA}

\author[0000-0002-6465-3639]{D. Michael Crenshaw}
\affiliation{Department of Physics and Astronomy, Georgia State University, 25 Park Place, Suite 605, Atlanta, GA 30303, USA}

\author[0000-0002-9946-4731]{Marc Rafelski}
\affiliation{Space Telescope Science Institute, 3700 San Martin Drive, Baltimore, MD 21218, USA}
\affiliation{Department of Physics and Astronomy, Johns Hopkins University, Baltimore, MD 21218, USA}

\author[0000-0002-6928-9848]{Steven B. Kraemer}
\affiliation{Institute for Astrophysics and Computational Sciences, Department of Physics, The Catholic University of America, Washington, DC 20064, USA}

\author[0000-0001-5862-2150]{Garrett E. Polack}
\affiliation{Department of Physics and Astronomy, Georgia State University, 25 Park Place, Suite 605, Atlanta, GA 30303, USA}

\author[0000-0001-8112-3464]{Anna Trindade Falc\~{a}o}
\affiliation{Institute for Astrophysics and Computational Sciences, Department of Physics, The Catholic University of America, Washington, DC 20064, USA}

\author[0000-0002-3365-8875]{Travis C. Fischer}
\affiliation{AURA for ESA, Space Telescope Science Institute, 3700 San Martin Drive, Baltimore, MD 21218, USA}

\author[0000-0001-8658-2723]{Beena Meena}
\affiliation{Department of Physics and Astronomy, Georgia State University, 25 Park Place, Suite 605, Atlanta, GA 30303, USA}

\author[0000-0001-5099-8700]{Francisco Martinez}
\affiliation{Center for Relativistic Astrophysics, School of Physics, Georgia Institute of Technology,
Atlanta, GA 30332, USA}

\author[0000-0003-2450-3246]{Henrique R. Schmitt}
\affiliation{Naval Research Laboratory, Washington, DC 20375, USA}

\author[0000-0002-8837-8803]{Nicholas R. Collins}
\affiliation{Telophase Corporation at NASA's Goddard Space Flight Center, Code 667, Greenbelt, MD 20771, USA}

\author[0000-0001-7238-7062]{Julia Falcone}
\affiliation{Department of Physics and Astronomy, Georgia State University, 25 Park Place, Suite 605, Atlanta, GA 30303, USA}

\begin{abstract}
Active galactic nuclei (AGN) can launch outflows of ionized gas that may influence galaxy evolution, and quantifying their full impact requires spatially resolved measurements of the gas masses, velocities, and radial extents. We previously reported these quantities for the ionized narrow-line region (NLR) outflows in six low-redshift AGN, where the gas velocities and extents were determined from Hubble Space Telescope long-slit spectroscopy. However, calculating the gas masses required multi-component photoionization models to account for radial variations in the gas densities, which span $\sim$6 orders of magnitude. In order to simplify this method for larger samples with less spectral coverage, we compare these gas masses with those calculated from techniques in the literature. First, we use a recombination equation with three different estimates for the radial density profiles. These include constant densities, those derived from [S~II], and power-law profiles based on constant values of the ionization parameter ($U$). Second, we use single-component photoionization models with power-law density profiles based on constant $U$, and allow $U$ to vary with radius based on the [O~III]/H$\beta$ ratios. We find that assuming a constant density of $n_\mathrm{H} =$~10$^2$~cm$^{-3}$ overestimates the gas masses for all six outflows, particularly at small radii where the outflow rates peak. The use of [S~II] marginally matches the total gas masses, but also overestimates at small radii. Overall, single-component photoionization models where $U$ varies with radius are able to best match the gas mass and outflow rate profiles when there are insufficient emission lines to construct detailed models.
\end{abstract}

 %--- galaxies: kinematics and dynamics

\section{Introduction}

\subsection{Feedback from Mass Outflows in Active Galaxies}

Mass outflows of ionized and molecular gas driven by active galactic nuclei (AGN) may regulate the coevolution of supermassive black holes (SMBHs) and their host galaxies by evacuating the bulge of star-forming gas \citep{Ciotti2001, Hopkins2005, Kormendy2013, Heckman2014, Fiore2017, Cresci2018, Harrison2018, Storchi-Bergmann2019, Veilleux2020, Florez2021, Laha2021}. Outflows connecting the sub-parsec central engine to the kiloparsec scale galaxy environment reside in the narrow-line region (NLR), consisting of ionized gas $\sim$1~$-$~1000+ parsecs (pcs) from the SMBH \citep{Osterbrock2006}. These outflows extend from the smallest scales that can be spatially-resolved in nearby galaxies to bulge-galaxy scales where they may affect star formation. The authors have an ongoing program to determine whether or not NLR outflows are providing significant feedback to their host galaxies by quantifying the outflowing masses ($M$) and velocities ($v$) in radial intervals ($\delta r$) as a function of distance from their SMBHs. These parameters are then used to calculate the outflow energetics, including mass outflow rates ($\dot M = M v / \delta r$), kinetic energies ($ E = 1/2 M v^2$), kinetic energy flow rates ($\dot E = 1/2 \dot M v^2$), momenta ($p = Mv$), and momenta flow rates ($\dot p = \dot M v$). We have completed this analysis for six nearby ($z \leq 0.05$) AGN, with the results presented in \cite{Revalski2021}, and the properties of the sample listed in Table~\ref{sample}.

Significant effort is being devoted to understanding the physics of AGN-driven outflows, and measuring their energetic impact (e.g. \citealp{Fischer2018, Baron2019a, Baron2019b, ForsterSchreiber2019, Mingozzi2019, Davies2020Rebecca, Wylezalek2020, Avery2021, Fluetsch2021, Lamperti2021, Luo2021, Negus2021, Ruschel-Dutra2021, Speranza2021, Vayner2021, Bianchin2022, Deconto-Machado2022, Kakkad2022}). However, outflows are inhomogeneous, with a large range of densities and geometries, such that the derived gas masses and outflow rates depend strongly on how the gas densities are estimated. This can result in outflow energetic estimates that differ by $\sim$1~$-$~3 dex for the same galaxies \citep{Karouzos2016, Bischetti2017, Perna2017}. Without a benchmark to compare with, it is unclear what the uncertainties and systematic biases are in each respective method. Therefore, it is critical to carefully examine and compare commonly used techniques for determining the outflow densities. This has been recently explored in several studies \citep{Baron2019b, Davies2020Ric}; however, there are only a small number that utilize high spatial resolution observations \citep{Dall'AgnoldeOliveira2021} and photoionization models \citep{Baron2019b, TrindadeFalcao2021}.

\begin{figure*}[ht!]
\centering
\includegraphics[width=0.49\textwidth]{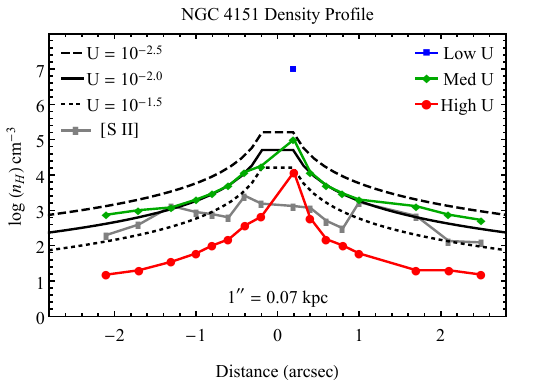}
\includegraphics[width=0.49\textwidth]{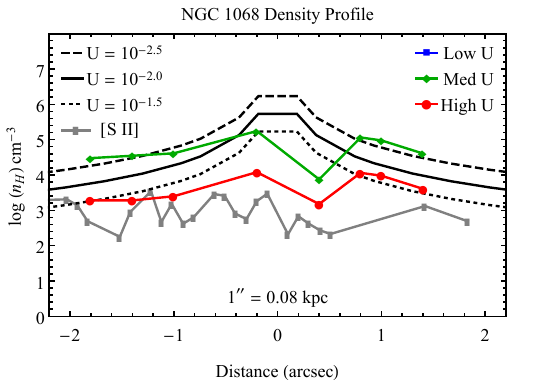}
\includegraphics[width=0.49\textwidth]{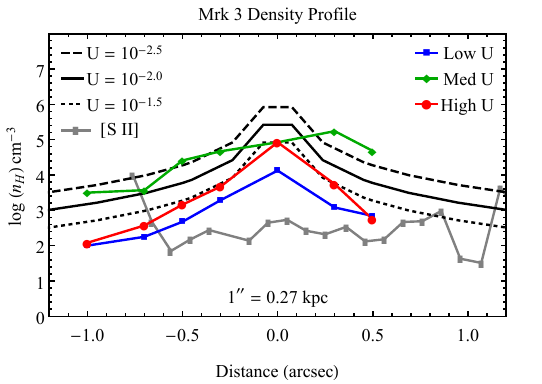}
\includegraphics[width=0.49\textwidth]{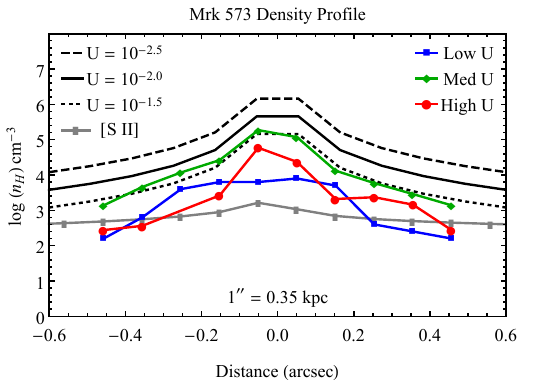}
\includegraphics[width=0.49\textwidth]{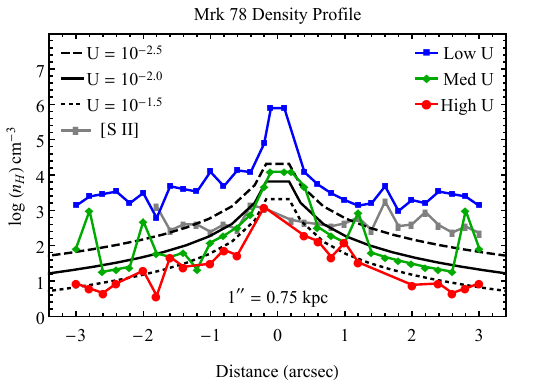}
\includegraphics[width=0.49\textwidth]{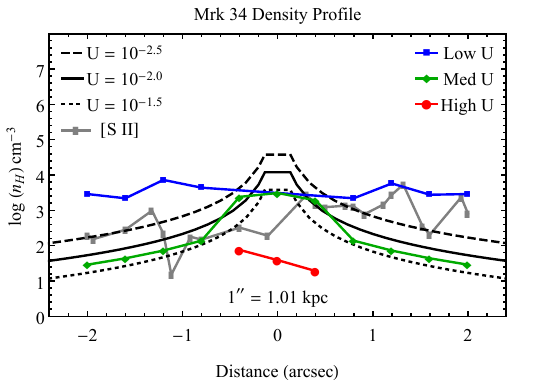}
\caption{The radial gas density profiles used in this study, compared to the densities from multi-component Cloudy models \citep{Revalski2021}. The red circles, green diamonds, and blue squares represent the high, medium, and low ionization Cloudy model components, respectively. Power-law density profiles that result from assuming constant ionization parameters with radius are shown by black lines, with values of log($U$) = $-$1.5 (dotted), $-$2.0 (solid), and $-$2.5 (dashed), where the vertical positions are set by the AGN luminosity (see Equation~\ref{uion}). Densities derived from the [S~II] emission line ratios are shown by gray rectangles. In general, lower ionization parameters correspond to higher densities, except for the low ionization components of Mrk~3 and Mrk~573. In those cases, the ionizing flux was reduced by absorption at smaller radii (see \citealp{Collins2009} and \citealp{Revalski2018b}), and their ionization parameters cannot be compared to the black curves. There is no low ionization component for NGC~1068, as the observations were well fit with two-component models. The [S~II] measurements for Mrk~573, and the model components for Mrk~34, are based on ground-based data that are susceptible to atmospheric blending at small radii. Finally, it is worth noting that the medium ionization model components generally follow log($U$)~$\approx$~$-$2.0 (density decreasing $\propto$~1/$r^2$), and produce the majority of the observed [O~III] emission. References for the photoionization models are: NGC~4151 \citep{Kraemer2000c}, NGC~1068 \citep{Kraemer2000b}, Mrk~3 \citep{Collins2009}, Mrk~573 \citep{Revalski2018a}, Mrk~78 \citep{Revalski2021}, and Mrk~34 \citep{Revalski2018b}. The uncertainties in the model densities are $\pm$0.05~dex, based on the model grid resolution. The data shown in this figure are tabulated in Appendix~\ref{appendix2a}.}
\label{densities}
\end{figure*}

In \cite{Revalski2021}, we found that spatially-resolved observations are essential for localizing AGN feedback and determining the most accurate outflow parameters. In addition, multi-component photoionization models were used to constrain the gas densities and precisely calculate the mass of the ionized outflows. As shown in Figure~\ref{densities}, we found that up to three photoionized gas components of different densities (and thus ionization parameters, hereafter $U$) were needed to match the multitude of emission lines in the high spatial resolution Hubble Space Telescope (HST) Space Telescope Imaging Spectrograph (STIS) long-slit spectra. In general, we found that the medium ionization component produces the majority of the [O~III] emission, and traces a significant portion of the ionized gas mass, accounting for $\sim$20~$-$~95\% of the mass in all components. The tenuous high ionization component can contribute significantly to the mass, particularly at small radii. In contrast, the dense low ionization component emits efficiently, contributing significantly to the observed luminosity with only a small fraction of the ionized gas mass \citep{Revalski2021, TrindadeFalcao2021}.

These models reveal that the gas densities vary by more than six orders of magnitude between different objects and components, and typically decrease in density as a function of distance from the nucleus. While these results provide the largest sample to date of spatially-resolved NLR mass outflow energetics based on multi-component photoionization models, generating the models is a time intensive and computationally expensive process. Most importantly, creating accurate models requires high signal-to-noise (S/N) and spatially resolved spectral observations of numerous emission lines for each AGN, which are rarely available.

\begin{deluxetable*}{lccccccccccc}
\vspace{0.5em}
\tablecaption{Physical Properties of the Active Galaxy Sample}
\tablehead{
\colhead{Catalog} & \colhead{Redshift} & \colhead{Distance} &\colhead{Scale} & \colhead{log(\lbol)} & \colhead{log(\mbh)} & \colhead{\ledd} & \colhead{log(Q(H))} & \colhead{[O~III]/H$\beta$} & \colhead{E(B-V)} & \colhead{References}\\[-0.5em]
\colhead{Name} & \colhead{(21 cm)} & \colhead{(Mpc)} &\colhead{(pc/$\arcsec$)} & \colhead{(erg s$^{-1}$)} & \colhead{($M_{\odot}$)} & \colhead{(ratio)} & \colhead{(phot s$^{-1}$)} & \colhead{(ratio)} & \colhead{(mag)} & \colhead{(Cols. 5, 6)}\\[-0.5em]
\colhead{(1)} & \colhead{(2)} & \colhead{(3)} &\colhead{(4)} & \colhead{(5)} & \colhead{(6)} & \colhead{(7)} & \colhead{(8)} & \colhead{(9)} & \colhead{(10)} & \colhead{(11)}
}
\startdata
NGC 4151 & 0.0033 & 13.3 & 67.4 & 43.9 & 7.6 & 0.01 & 53.5 & 12 & 0.18 & 1, 2 \\
NGC 1068 & 0.0038 & 16.0 & 77.6 & 45.0 & 7.2 & 0.50 & 54.6 & 15 & 0.38 & 3, 3 \\
Mrk 3    & 0.0135 & 56.6 & 274.5 & 45.3 & 8.7 & 0.04 & 54.6 & 13 & 0.22 & 4, 3 \\
Mrk 573  & 0.0172 & 72.0 & 349.1 & 45.5 & 7.3 & 0.75 & 54.8 & 13 & 0.28 & 5, 3 \\
Mrk 78   & 0.0372 & 154.2 & 747.4 & 45.9 & 7.9 & 0.79 & 54.6 & 12 & 0.40 & 6, 6 \\
Mrk 34   & 0.0505 & 207.9 & 1007.7 & 46.2 & 7.5 & 3.98 & 54.9 & 11 & 0.30 & 7, 8 \\
\enddata
\tablecomments{Columns are (1) target name, (2) 21~cm redshift from the NASA/IPAC Extragalactic Database, (3) Hubble distance and (4) spatial scale assuming $H_0$ = 71 km s$^{-1}$ Mpc$^{-1}$, (5) bolometric luminosity, (6) black hole mass, (7) the corresponding Eddington ratio (\ledd) calculated using $L_{Edd} = 1.26 \times 10^{38} \left(M/M_{\odot}\right)$ erg~s$^{-1}$, (8) the number of ionizing photons per second emitted by the AGN, (9) the mean [O~III]/H$\beta$ ratio across the NLR, and (10) the mean color excess E(B-V) across the NLR. Column (11) gives the references for columns~5~$-$~6, which are: (1)~\citealp{Crenshaw2012}, (2)~\citealp{Bentz2006}, (3)~\citealp{Woo2002}, (4)~\citealp{Collins2009}, (5)~\citealp{Revalski2018a}, (6)~\citealp{Revalski2021}, (7)~\citealp{Revalski2018b}, and (8)~\citealp{Oh2011}.}
\label{sample}
\vspace{-2em}
\end{deluxetable*}

\subsection{The Search for a Simplified Methodology}

Using the results of \cite{Revalski2021} as a benchmark, we now explore methods of mass determination that have less stringent data requirements, with the goal of developing a more streamlined process that can be applied to large samples of spatially resolved outflow observations\footnote{A portion of this analysis has been adapted from \cite{Revalski2019a}: ``Quantifying Feedback from Narrow Line Region Outflows in Nearby Active Galaxies", Dissertation, Georgia State University, 2019. \url{https://scholarworks.gsu.edu/phy_astr_diss/114}.}. These comparison tests will reveal the effects of various physical assumptions on the resulting mass estimates, and allow us to determine if a simplified method that has less demanding data and modeling requirements may be implemented to determine accurate ionized gas masses and outflow rates. For example, there are several dozen AGN with high quality archival HST STIS spectroscopy utilizing the G430M grating. This medium dispersion grating allows for precise radial velocity measurements, but the narrow spectral range only encompasses the [O~III] and in some cases also the H$\beta$ emission lines. These types of data would not be suitable for detailed photoionization modeling, as there are insufficient emission lines to constrain the physical conditions in the gas. However, if a sufficiently accurate simplified technique based on the luminosity of a single line can be found, then the number of AGN that we may derive high spatial resolution gas mass and outflow rate profiles for increases significantly. Without the results from multi-component photoionization models \citep{Revalski2021}, we would be unable to estimate the reliability and systematic uncertainties of these techniques.

The results of many outflow studies rest upon the ability to accurately determine the mass of ionized gas that is responsible for producing the observed emission line luminosities. In \cite{Revalski2021}, we generated spatially resolved, multi-component photoionization models to match the relative emission line strengths and reveal the physical conditions in the gas such as number density, column density, and ionization state. These models capture all of the physical processes that lead to photon emission, and allow us to predict the luminosity per unit mass that is emitted by the ionized gas.

The physical conditions in the gas can vary significantly on small scales that may be unresolved even in high spatial resolution HST STIS observations, as evidenced by the wide range of ionized species present in the extracted spectra. The $\sim$0\farcs1 spatial resolution of HST STIS corresponds to physical scales of $\sim$5~$-$~100~pc for the AGN in this study. Any variations of the physical conditions on smaller scales are blended into the single spectrum extracted at each location along the slit. Multi-component photoionization models account for this by allowing for multiple density and ionization states, with the predicted emission from all components convolved to match the observed spectrum, and represent the best determination of ionized gas masses with the current generation of telescopes and instruments \citep{Groves2004, Netzer2013}.

In our earlier studies, we briefly explored mass estimates derived from geometric (e.g. \citealp{Muller-Sanchez2011}) and luminosity-based (e.g. \citealp{Bae2017}) techniques, which revealed that geometric techniques (e.g., assuming a geometry and filling factor) can be subject to strong systematic biases, and that luminosity-based techniques that employ a physical tracer of the gas mass produce more promising results (see \S8.1 of \citealp{Revalski2018a} for a more detailed discussion). For this reason, we explore mass estimates from luminosity-based techniques that may employ spectra with fewer emission lines, and potentially without the need to generate multi-component photoionization models for each AGN.

In \S2 we describe the recombination and photoionization equations used to calculate the gas masses and outflow rates, with the results presented in \S3. We discuss the implications of the results in \S4, and present our conclusions in \S5.

\textcolor{white}{.}

\section{Methods}
\label{sec:methods}

We assume that the observer has spatially resolved spectroscopy of a bright emission line such as [O~III]~$\lambda$5007~\AA~to determine the velocity and extent of the outflowing ionized gas, and an emission-line image to determine the luminosity of the gas outside of the spectral slit \citep{Revalski2021}. Alternatively, integral field unit (IFU) observations of the entire NLR at sufficiently high ($\sim$0\farcs1) spatial resolution would fulfill these requirements. In this section, we describe the two equations that will be used to calculate gas masses.

\subsection{Gas Masses from Pure Recombination}

The first technique that we consider uses a simple recombination equation to estimate the gas masses. This process converts an emission line luminosity to gas mass by assuming each photon originated from simple radiative recombination, which allows the luminosity and mass to be related by a constant, known as the recombination coefficient, at a given density. We provide a detailed derivation in Appendix~\ref{appendix1}, which yields that the H$\beta$~$\lambda$4861~\AA~emission line luminosity ($L_{\mathrm{H}\beta}$) and the ionized gas mass ($M_{ion}$) are related by
\begin{equation}\label{msimp}
M_{ion} = (12.48~_{-3.83}^{+3.70}\times10^{-33})\left(\frac{L_{\mathrm{H}\beta}}{n_e}\right)~\mathrm{(M_\odot)}.
\end{equation}
This expression is conceptually appealing, as it only requires a measurement of the emission line luminosity and electron density to determine the gas mass within a region. However, there are several observational considerations that add an underlying level of complexity, as discussed in the Appendix.

Using this expression, the remaining free parameter is the gas density, and in \S\ref{sec:results} we will explore 1)~assuming a constant value for the density across the NLR, 2)~using densities derived from the [S~II] doublet, and 3)~using a power-law density profile assuming a constant $U(r)$ across the NLR.

\subsection{Gas Masses from Photoionization Models}

The second technique that we consider is a simplification of our multi-component photoionization modeling process \citep{Kraemer2000b, Crenshaw2015, Revalski2021}. There, we calculated the masses for each of the high, medium, and low ionization components (Figure~\ref{densities}) separately by dividing up the H$\beta$ luminosity by the model fractional contributions, and then summed the component masses. In this case, we use a single-component, optically-thick photoionization model to calculate the ionized gas mass at each radius, which is given by
\begin{equation}
M_{ion} = N_\mathrm{H} \mu m_p \left(\frac{L_{\mathrm{H}\beta}}{F_{\mathrm{H}\beta_m}}\right),
\label{masseqn}
\end{equation}
\noindent
where $N_\mathrm{H}$ is the model hydrogen column density, $\mu =$~1.4 is the mean mass per proton, $m_p$ is the proton mass, $F_{\mathrm{H}\beta_m}$ is the H$\beta$ model flux, and $L_{\mathrm{H}\beta}$ is the luminosity of H$\beta$ calculated from the extinction-corrected flux and distance\footnote{Conceptually, the ratio of the luminosities and model fluxes is the area of the clouds, which are multiplied by the column densities (projected particles per unit area). This provides the total number of particles, which is multiplied by the mean mass per particle to give the ionized gas mass at each radius.}. In this case, the gas density (n$_\mathrm{H}$) is an input parameter for the photoionization models, which yield $N_\mathrm{H}$ and $F_{\mathrm{H}\beta_m}$.

Using this expression in \S\ref{sec:results}, we will explore power-law density profiles based on 1)~a constant $U(r)$ and 2)~a variable $U(r)$ across the NLR. With spatially-resolved observations, where the distance of the emitting gas from the AGN is known, it is possible to determine the hydrogen number density ($n_\mathrm{H}$) explicitly for a given $U$ from the ionization parameter equation. Specifically,
\begin{equation}
n_\mathrm{H} = \left(\frac{Q(H)_{ion}}{4 \pi r^2 c~U}\right)
\label{uion}
\end{equation}
\noindent
where $Q(H)_{ion}$ is the number of ionizing photons s$^{-1}$ emitted by the AGN, assuming no absorption by intervening gas at smaller radii, $r$ is the radial distance of the gas from the AGN, $U$ is the ionization parameter, which is the ratio of the number of ionizing photons to hydrogen atoms at the face of the gas cloud, and $c$ is the speed of light. With an estimate or determination of $U$ from the spectroscopy, and a value of $Q(H)_{ion}$ from the AGN continuum luminosity and integrating the spectral energy distribution (SED), the density $n_\mathrm{H}$ can be obtained at each radial position.

As discussed by several authors (e.g. \citealp{Kraemer2000b, Davies2020Ric}), we may expect the [O~III] emission to arise in gas with log($U$) $\approx$~$-$2 at each radius, because the [O~III]/H$\beta$ ratios are generally $\sim$10~$-$~20, and constant across the NLR. The latter requires that the density drop $\propto 1/r^2$ as the ionizing flux does the same, maintaining constant ionization. This process was used by \cite{TrindadeFalcao2021} to estimate the ionized gas masses and outflows rates for a sample of QSO2s. However, using this density profile, it is not clear what total fraction of the ionized gas is traced out by the [O~III] luminosity, and how this may change as a function of radius, as the NLR can have a range of clouds with different $U$ at each position \citep{Kraemer2000b, Collins2009}. We are now in a unique position to explore this, by testing the assumptions of other techniques through comparison with the multi-component modeling results \citep{Revalski2021}.

\subsection{Calculations}

Independent of the various methods, all luminosity-based techniques require three fundamental parameters to estimate the ionized gas masses and outflow rates. These are radial profiles (or global estimates) of 1)~the emission line luminosity, 2)~the outflow velocity, and 3)~the outflow gas density. With these observational considerations, we may adopt the emission line luminosities and deprojected radial outflow kinematics from our earlier studies \citep{Crenshaw2015, Revalski2018a, Revalski2018b, Revalski2021} as control variables, because these may be determined with equal precision from moderate dispersion spectra of a single emission line. The remaining element is to determine an accurate density law profile, calculate the gas masses with the recombination (Equation~\ref{msimp}) and photoionization (Equation~\ref{masseqn}) relations, and determine which technique best reproduces the multi-component results.

We consider the cases listed above as they are commonly employed in the literature and/or have the most relaxed data requirements. In all cases, the H$\beta$ luminosity is initially approximated by using the [O~III] luminosities derived from our [O~III] imaging, and scaling them by the mean reddening-corrected [O~III]/H$\beta$ ratio for each galaxy. While a marginal increase in accuracy of estimating the H$\beta$ flux may be obtainable by using the exact [O~III]/H$\beta$ ratio at each location along the slit, the range of [O~III]/H$\beta$ ratios across the NLR are approximately constant, and are provided in Table~\ref{sample}. These averages are accurate to within a factor of two at all modeled radii for all six galaxies, except Mrk~78 where a factor of $\sim$3 change is seen at large radii. This further relaxes the requirement that a spectrum used with a simplified method need have H$\beta$ present at all locations; however, we will revisit the benefits of having both lines later. In this framework, the extinction-corrected H$\beta$ luminosity at each radius is given by
\begin{equation}
L_{\mathrm{H}\beta} = \left(4\pi D^2 F_{\mathrm{[O~III]}}\right) \times \left(\frac{F_{\mathrm{H\beta}}}{F_{\mathrm{[O~III]}}} \right) \times \left(10^{0.4 \cdot R_{\lambda 5007} \cdot E(B-V)}\right)
\end{equation}
where $D$ is the distance to the galaxy assuming all portions of the NLR are at approximately the same distance from us, $F_{\mathrm{[O~III]}}$ is the measured [O~III] $\lambda$5007~\AA~flux from our HST images, $R_{\lambda 5007} = 3.57$ is the value of the Galactic reddening curve for [O~III] $\lambda$5007~\AA, and E(B-V) is the color excess from extinction given by
\begin{equation}\label{eqn:ebmv}
E(B-V) = -\frac{2.5\log\left(\frac{F_{o}}{F_{i}}\right)}{R_\lambda} = \frac{2.5\log\left(\frac{(\mathrm{H}\alpha/\mathrm{H}\beta)_{i}}{(\mathrm{H}\alpha/\mathrm{H}\beta)_{o}}\right)}{R_{\mathrm{H}\alpha}-R_{\mathrm{H}\beta}}
\end{equation}
where $F_o$ and $F_i$ are the observed and intrinsic fluxes, respectively. The flux ratios can be expanded to the intrinsic and observed H$\alpha$/H$\beta$ ratios, and the Galactic reddening values are $R_{\mathrm{H}\alpha} \approx$ 2.497 and $R_{\mathrm{H}\beta} \approx$ 3.687 \citep{Savage1979, Cardelli1989}, assuming the standard Galactic reddening law applies within both our Galaxy and within the AGN host galaxy. For each target, the mean color-excess was calculated from the spatially resolved values \citep{Crenshaw2015, Revalski2018a, Revalski2018b, Revalski2021}, and are provided in Table~\ref{sample}. In cases where there are not two or more recombination lines to estimate the extinction, a global value from ground-based observations or other measurements would be required, introducing further uncertainty.

\subsection{Sources of Uncertainty}

Before exploring the results, it is worth delineating the uncertainties that are inherent to both our primary method and these simplified techniques. These include errors in the distances to the galaxies, the geometric models for deprojecting velocities and distances from the SMBH, and the choice of reddening curve. These specific issues are addressed in \S7.1 of \cite{Revalski2021}. A critical concern for this study is related to the bolometric luminosities. Specifically, systematic errors in the adopted luminosities (Table~\ref{sample}) will directly affect the densities calculated from Equation~\ref{uion}, for methods based on the ionization parameter. In most cases, the bolometric luminosities are based on the [O~III] emission line, and were confirmed with X-ray models or continuum estimates. Specifically, the luminosity for NGC~4151 was estimated from the 5100~\AA~continuum \citep{Bentz2006, Crenshaw2012}, while Mrk~573 is based on an infrared to X-ray luminosity conversion \citep{Melendez2008, Kraemer2009}. Similarly, the value for Mrk~34 is based on both [O~III] and Nuclear Spectroscopic Telescope Array (NuSTAR) 3~$-$~40~keV observations \citep{Gandhi2014}. The adopted luminosity for Mrk~78 is higher than some literature values \citep{Woo2002}, but is consistent with the very extended NLR, and Chandra X-ray modeling (Maksym et al., \textit{private communication}). These consistency checks are vital because the [O~III] to bolometric luminosity scaling relationship has an uncertainty of $\pm$0.4 dex \citep{Heckman2004}, which could dominate the uncertainties in Equation~\ref{uion}. The AGN luminosities can also vary on short timescales, so the light-travel time average across the NLR provides a valuable constraint on the ionization history. Most importantly, self-consistent luminosities have been used for all methods for each AGN in this study.

\section{Results}
\label{sec:results}

We consider the five cases described in \S\ref{sec:methods}, which each estimate the gas densities and masses based on different physical assumptions. We then use these gas masses ($M_{ion}$), together with the deprojected radial outflow velocities ($v$) and spatial bin sizes ($\delta r$), to calculate the mass outflow rates ($\dot M_{ion}$ = $M v / \delta r$). We present the resulting ionized gas masses and outflow rates on logarithmic scales in Figures~\ref{constrecomb}, \ref{s2recomb}, \ref{uionrecomb}, and~\ref{uioncloudy}.

\subsection{Recombination with Constant Densities}\label{sec:const}

The first method that we consider is to fix the density at a constant value across the entire NLR, and calculate the gas masses from the recombination relation (Equation~\ref{msimp}). When density diagnostics are not available from spectra, this technique has often been employed adopting a density of $n_\mathrm{H} =$~10$^2$~cm$^{-3}$ \citep{Kakkad2018, Nevin2018}, although a case for higher average densities has been made more recently \citep{Baron2019b, Davies2020Ric, Kakkad2020}. We examine constant density values of $n_\mathrm{H} =$~10$^2$, 10$^3$, and 10$^4$ cm$^{-3}$, as this range encompasses that observed for the majority of our photoionization model components (Figure~\ref{densities}). In examining Equation~\ref{msimp}, it is worth noting that a lower density corresponds to a higher mass for a fixed value of the luminosity.

The results of this process are shown in Figure~\ref{constrecomb}, where we find that adopting a constant density of $n_\mathrm{H} =$~10$^2$~cm$^{-3}$ overestimates the gas mass for all six galaxies by factors ranging from approximately 2~$-$~60. Adopting a higher density of $n_\mathrm{H} =$~10$^3$ cm$^{-3}$ still overestimates the mass for NGC~4151 and NGC~1068. Mrk~573 must be considered separately for $r > 175$ pc, as [S~II] densities were used at larger radii due to a lack of emission lines (see \citealp{Revalski2018a}), but at smaller radii the mass is also overpredicted for $n_\mathrm{H} =$~10$^3$ cm$^{-3}$. While assuming a constant density is able to reproduce the total ionized gas masses for Mrk~3, Mrk~78, and Mrk~34 to within factors of approximately 2~$-$~4, there are deviations greater than 1 dex at some radii, resulting in larger errors on the maximum mass outflow rates.

These results are consistent with the physical model that the higher density material at small radii, which has a much higher emissivity, is contributing a significant fraction of the luminosity with a relatively small fraction of the mass, as $M_{ion} \propto L/n_\mathrm{H}$. The simplifying assumption that the density is constant with radius significantly alters the mass and mass outflow rate profiles, showing that the commonly adopted value of $n_\mathrm{H} =$~10$^2$~cm$^{-3}$ is inaccurate for the NLR of every object in the sample. While assuming a higher constant value of $n_\mathrm{H}\approx$~10$^3$~cm$^{-3}$ is an improvement, it can still lead to large errors at specific locations in the NLR for each AGN.

\begin{figure*}[ht!]
\vspace{1.0em}
\centering
\includegraphics[scale=0.97]{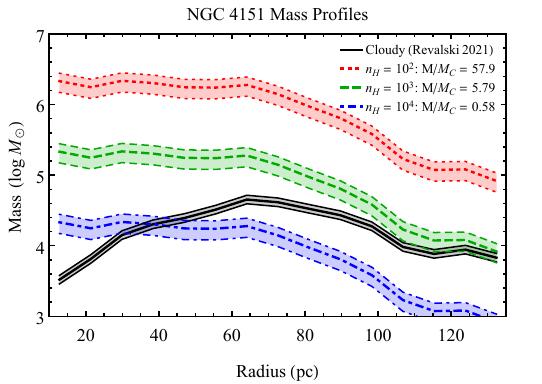}
\includegraphics[scale=0.97]{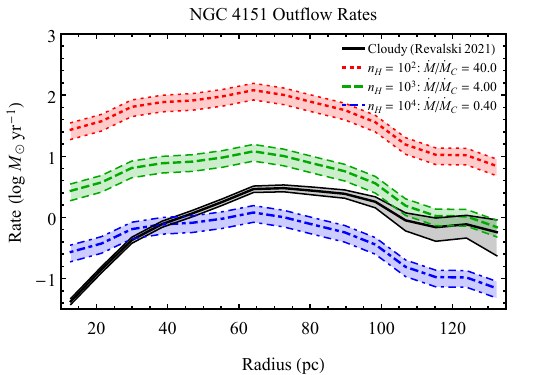}\\\vspace{1.0em}
\includegraphics[scale=0.97]{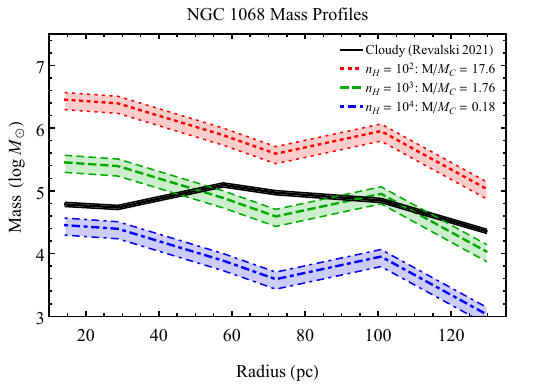}
\includegraphics[scale=0.97]{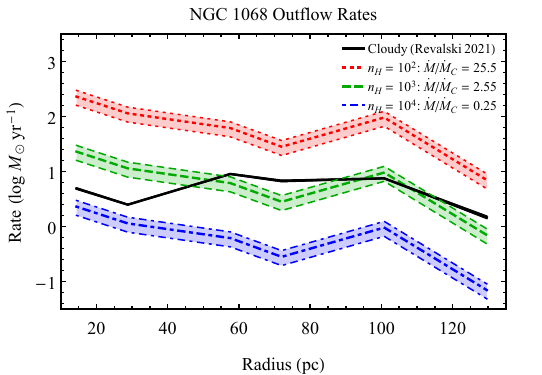}\\\vspace{1.0em}
\includegraphics[scale=0.97]{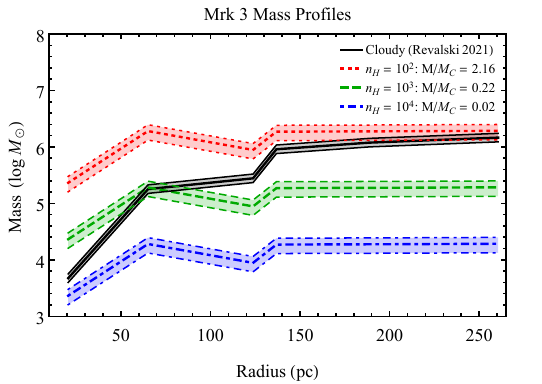}
\includegraphics[scale=0.97]{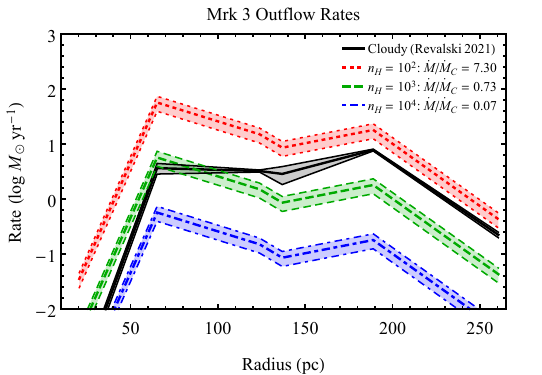}
\caption{The ionized gas mass (left column) and outflow rate (right column) profiles for NGC~4151, NGC~1068, Mrk~3, Mrk~573, Mrk~78, and Mrk~34, calculated from Cloudy models in solid black (\citealp{Revalski2021}), and assuming constant density profiles with radius of $n_{\mathrm{H}} =$~10$^2$ cm$^{-3}$ (dotted red), $n_{\mathrm{H}} =$~10$^3$ cm$^{-3}$ (dashed green), and $n_{\mathrm{H}} =$~10$^4$ cm$^{-3}$ (dashed-dotted blue) with the recombination relation (Equation~\ref{msimp}). The legend indicates the linear ratio of the total constant density mass to the Cloudy derived mass ($M/M_C$) and peak outflow rates ($\dot M/ \dot M_C$). The uncertainties in the simplified method are dominated by the uncertainty in $\alpha_{_{H\beta}}^{eff}$ due to its dependence on temperature. The largest deviations between the constant density and Cloudy model results are seen at small radii, where the commonly used values of $n_{\mathrm{H}} =$~10$^2$~$-$~10$^3$ cm$^{-3}$ significantly overestimate the gas masses and outflow rates. While assuming a constant density can marginally reproduce the total gas mass for the higher luminosity targets, the disagreement is a strong function of radial distance and exceeds 1~dex at one or more locations for all AGN when adopting $n_{\mathrm{H}} = 10^2$ cm$^{-3}$.}
\label{constrecomb}
\vspace{1.0em}
\end{figure*}
\addtocounter{figure}{-1}
\begin{figure*}[ht!]
\vspace{1.0em}
\centering
\includegraphics[scale=0.97]{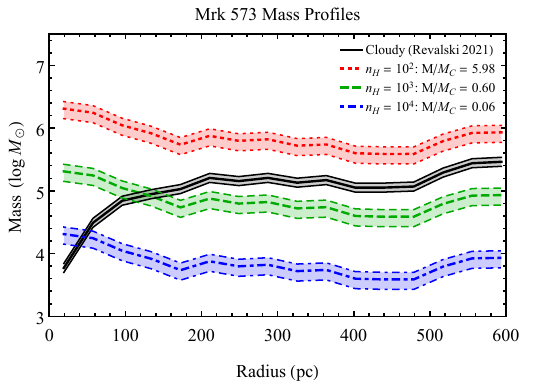}
\includegraphics[scale=0.97]{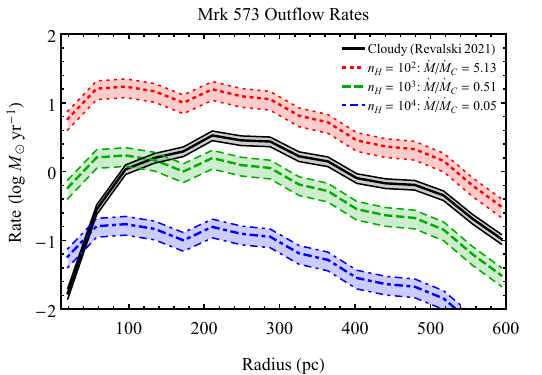}\\\vspace{1.0em}
\includegraphics[scale=0.97]{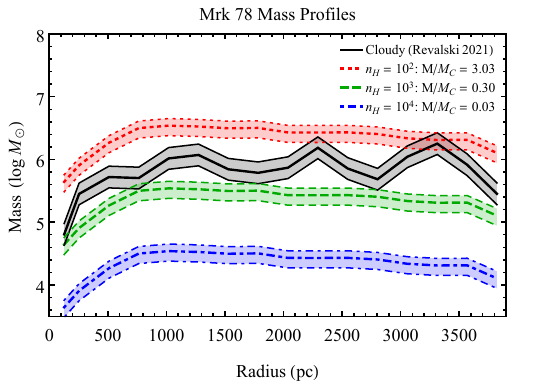}
\includegraphics[scale=0.97]{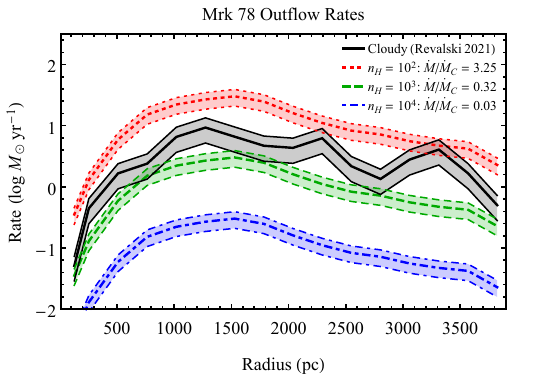}\\\vspace{1.0em}
\includegraphics[scale=0.97]{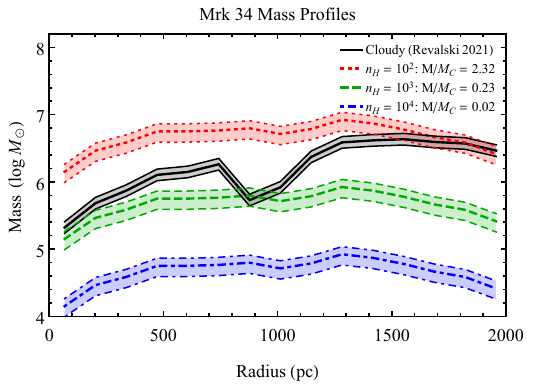}
\includegraphics[scale=0.97]{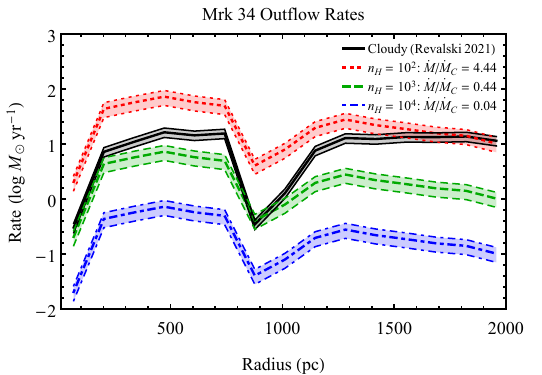}
\caption{{\it continued.}}
\vspace{1.0em}
\end{figure*}

Finally, it is worth noting that using the same constant density for a sample of AGN will introduce an artificial correlation with the bolometric luminosity. This can be seen from Equation~\ref{msimp}, where fixing the density with increasing luminosity will always result in larger gas masses. This assumption is only valid if the density of NLR outflows is independent of distance from the nucleus, as well as the AGN luminosity. Based on the photoionization modeling results shown in Figure~\ref{densities}, this is not a valid assumption, and adopting a constant value for the density is not recommended.

\subsection{Recombination with {\normalfont [S~II]} Densities}\label{sec:s2}

\begin{figure}%[hb!]
\centering
\includegraphics[width=0.49\textwidth]{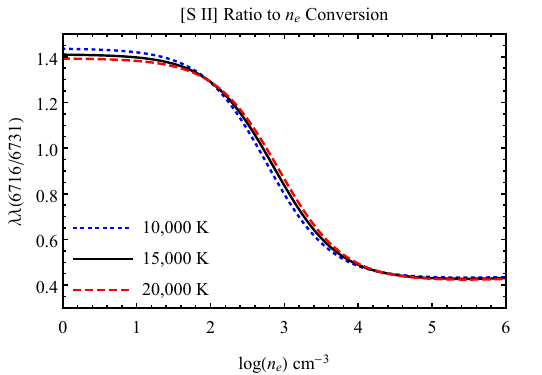}
\caption{The theoretical relationship between the [S~II] doublet emission line ratio and the electron density for common NLR temperatures \citep{Osterbrock2006}. These curves were calculated using the grid of Cloudy models presented in \cite{Revalski2018a}, and the data shown in this figure are tabulated in Appendix~\ref{appendix2b}.}
\label{tempdengrid}
\vspace{-0.5em}
\end{figure}

The second method that we consider is to derive the gas densities directly from density sensitive emission lines in the spectra, and calculate the gas masses from the recombination relation (Equation~\ref{msimp}). While there are several density sensitive doublets in the optical, such as [O~II] $\lambda \lambda$3726, 3729, [Ar~IV] $\lambda \lambda$4711, 4740, and [S~II] $\lambda \lambda$6716, 6731, only the [S~II] doublet is typically spectrally resolved and sufficiently bright to be useful as a diagnostic tool. However, as discussed by several authors (e.g. \citealp{Kraemer2000c, Davies2020Ric, Revalski2021}), the [S~II] lines only probe a single ionization state of the emission line gas that is cooler and often at a different density than the [O~III] emission lines that are used to trace the gas luminosity and kinematics. Additionally, the [S~II] doublet is only sensitive over the range of $n_e \approx$~10$^2$~$-$~10$^4$~cm$^{-3}$, and is a weak function of temperature, as shown in Figure~\ref{tempdengrid}. Also, a significant fraction of the [S~II] emission can arise from the partially-ionized zone in a NLR cloud, where the reduced free electron density can lead to an underestimate of the total hydrogen density (\citealp{Kraemer2000c, Davies2020Ric, Riffel2021}), or from gas outside of the bicone, which is ionized by partially absorbed radiation \citep{Collins2009}. Comparing the gas masses derived using densities from [S~II] to the multi-component models will allow us to quantify the degree to which these limitations affect the estimation of mass outflow parameters.

The electron density profiles were calculated using the reddening-corrected [S~II] $\lambda \lambda$6716, 6731 emission line ratios, which agree with the densities derived from the observed line ratios to within $\sim$1\% for typical levels of extinction in the NLR, and were converted to hydrogen densities using $n_\mathrm{H} \approx 0.85n_e$ \citep{Crenshaw2015}. These values were obtained from \cite{Kraemer2000a} Tables 1A - 1B for NGC~4151, \cite{Revalski2018a} for Mrk~573, \cite{Revalski2018b} for Mrk~34, and \cite{Revalski2021} for Mrk~78, respectively. In the cases of Mrk~3 and NGC~1068, we fit the [S~II] lines in the same spectra that were used by \cite{Collins2009} and \cite{Kraemer2000b}, respectively, following the [O~III] template method to isolate the kinematic components that we used in our previous studies \citep{Revalski2021}. In \cite{Revalski2018a}, we generated a fine grid of Cloudy photoionization models across a wide range of densities for several typical NLR temperatures, and recorded the predicted [S~II] emission line ratios for use as a reference table. We used this process to match each [S~II] ratio with the corresponding density, and to determine the uncertainties that are introduced by a range of temperatures. These grids were originally presented in \cite{Revalski2018a}, and are shown in Figure~\ref{tempdengrid} for clarity. In the cases of NGC~4151 and Mrk~34, the [S~II] doublet was not detected for a small number of spatial extractions corresponding to an [O~III] image flux measurement due to low S/N in the spectra. At these locations, we interpolated the density using the measured values at the next inner and outer radial distances.

The resulting density profiles are shown in Figure~\ref{densities}. Overall, the [S~II] densities tend to be lower at small radii than predicted by the photoionization models. In general, we would expect the low ionization gas that gives rise to the [S~II] emission to be at a higher density compared to the higher ionization gas that is at approximately the same distance from the SMBH, but that does not need to be the case when shielding is considered. In that case, the low ionization gas sees a reduced ionizing flux due to absorption at smaller radii, which allows for lower densities at a given ionization parameter (see Equation~\ref{uion}). This is the case for the low ionization model component in Mrk~573 \citep{Revalski2018a}, and all of the components in Mrk~3 \citep{Collins2009}. Under those circumstances, the low ionization component can be at a lower density than the higher ionization components.

The [S~II] derived densities were used with Equation~\ref{msimp} to calculate the radial gas mass and outflow rates for each galaxy, and the results are shown in Figure~\ref{s2recomb}. Prior to interpreting these results, it is important to note that Mrk~573 is a special case, as we adopted the [S~II] densities for $r >$~175~pc because there were insufficient emission lines in the HST spectra to create detailed photoionization models. Thus there is an expected agreement between the benchmark and simplified results with regard to overall shape, with a constant offset capturing the difference between using a model derived scale factor versus the recombination coefficient in Equation~\ref{msimp}.

\begin{figure*}[ht!]
\vspace{1.0em}
\centering
\includegraphics[scale=0.97]{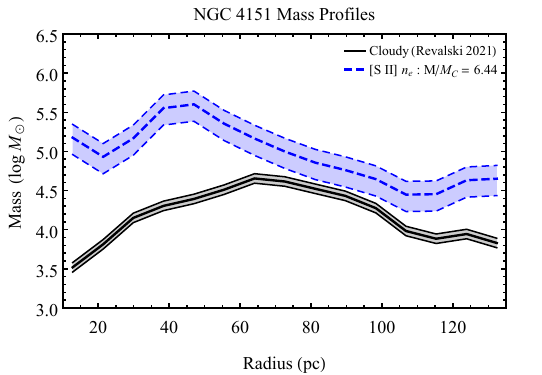}
\includegraphics[scale=0.97]{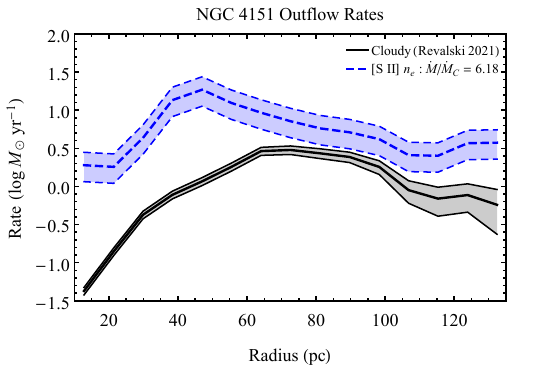}\vspace{1.0em}
\includegraphics[scale=0.97]{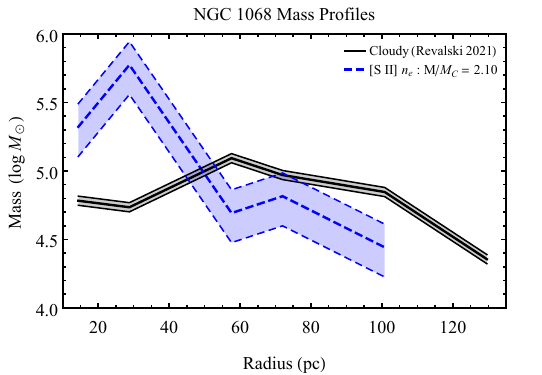}
\includegraphics[scale=0.97]{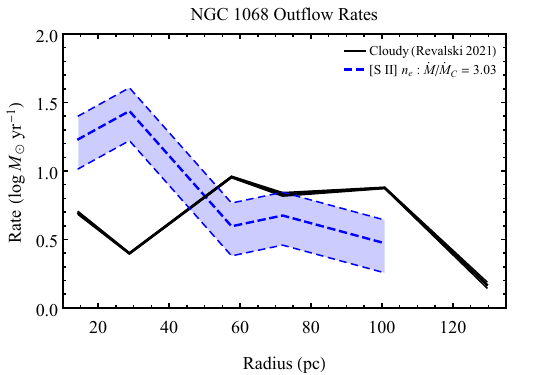}\vspace{1.0em}
\includegraphics[scale=0.97]{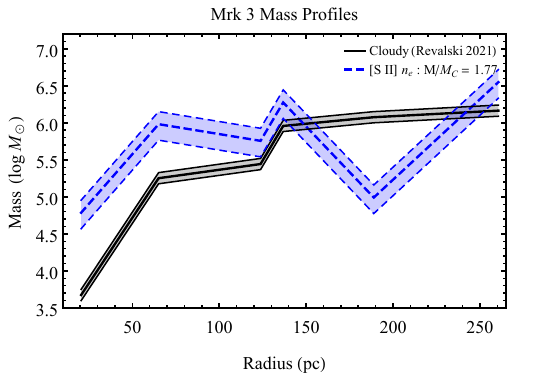}
\includegraphics[scale=0.97]{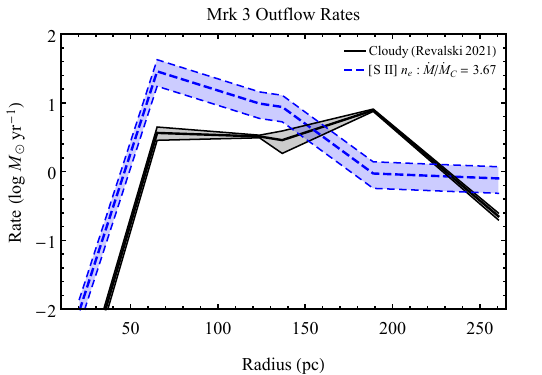}
\caption{The ionized gas mass (left column) and outflow rate (right column) profiles for NGC~4151, NGC~1068, Mrk~3, Mrk~573, Mrk~78, and Mrk~34, calculated from Cloudy models in solid black (\citealp{Revalski2021}), and using the density profiles derived from the [S~II] line ratios with the recombination relation (Equation~\ref{msimp}). Regions where the mass is overestimated correspond to underestimates in the gas density. The precise agreement in profile shape for Mrk~573 at $r >$~175~pc is expected as described in the text. The legend indicates the linear ratio of the total [S~II] to Cloudy derived mass ($M/M_C$) and peak outflow rates ($\dot M/ \dot M_C$). The uncertainties in the simplified method are dominated by the uncertainty in $\alpha_{_{H\beta}}^{eff}$ due to its dependence on temperature, and adopting a uniform uncertainty of 30\% in the [S~II] line ratios. This technique generally overestimates the gas masses and outflow rates at small radii, with the disagreement exceeding 1~dex at one or more radii for most of the galaxies.}
\vspace{4.0em}
\label{s2recomb}
\end{figure*}
\addtocounter{figure}{-1}
\begin{figure*}[ht!]
\vspace{0.5em}
\centering
\includegraphics[scale=0.97]{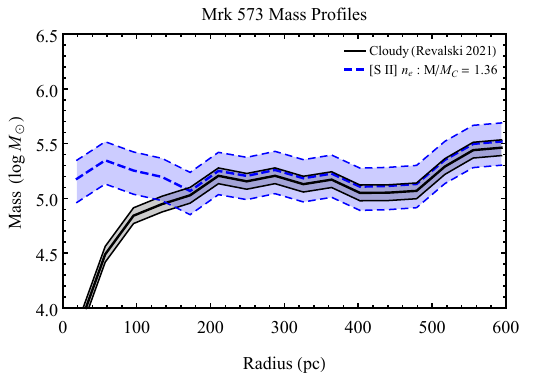}
\includegraphics[scale=0.97]{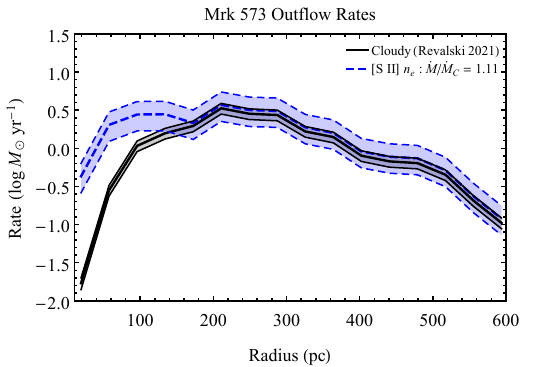}\vspace{0.5em}
\includegraphics[scale=0.97]{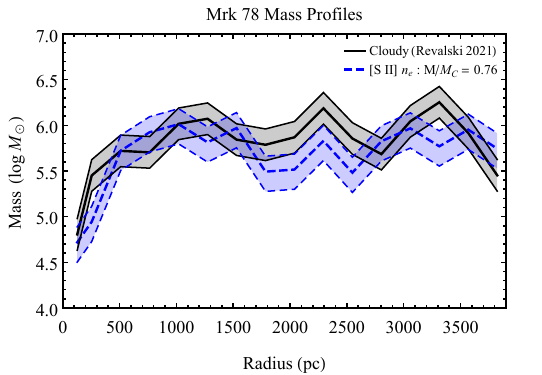}
\includegraphics[scale=0.97]{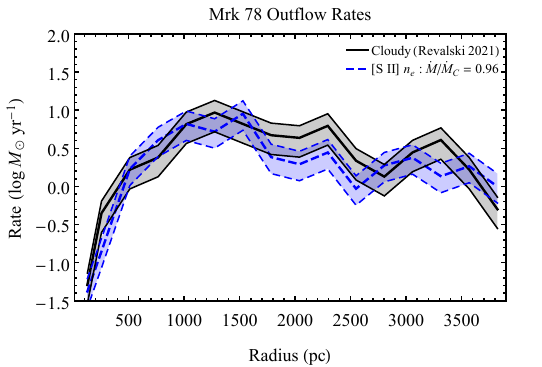}\vspace{0.5em}
\includegraphics[scale=0.97]{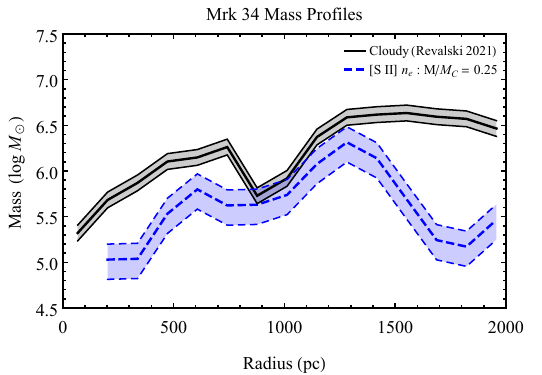}
\includegraphics[scale=0.97]{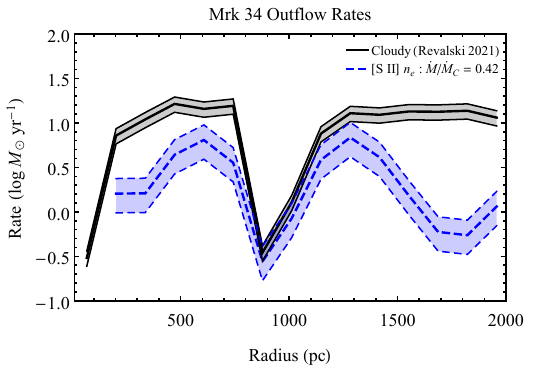}
\vspace{-1.0em}
\caption{{\it continued.}}
%\vspace{1.0em}
\end{figure*}

As shown in Figure~\ref{s2recomb}, this technique generally overestimates the ionized gas masses and outflow rates at small radii, with some exceptions. This is because the [S~II] line ratio will generally underestimate the overall density, and thus overestimate the gas mass, as they are inversely proportional. Physically, this is due to the denser material emitting more efficiently ($\propto n_e^2$, see Equation~\ref{emiss}), and thus dominating the line luminosity while only encompassing a small fraction of the total mass.

Surprisingly, this technique is able to reproduce the total ionized gas masses and peak outflow rates to within a factor of approximately six in all cases. However, it is worth highlighting that the differences at individual radii can be much larger, and exceed 1 dex in multiple cases. In addition, for clarity this analysis has adopted a typical 30\% formal uncertainty in the [S~II] line ratios based on average S/N and deblending, which were used to determine the density at each radius. For this reason, unless high S/N observations are available, the photoionization model densities are generally more tightly constrained, particularly at small radii where the gas densities exceed the density-sensitive range of the [S~II] doublet.

\subsection{Recombination with Constant U(r)}\label{sec:urecom}

The third method that we consider is to assume that the ionization parameter of the emission line gas is constant with radius such that $U(r) = $ constant. This will naturally lead to a density profile that decreases $\propto 1/r^{-log(U)}$ with distance from the nucleus if the ionizing luminosity is relatively constant over the light-travel time across the NLR, which is consistent with the results of our photoionization models for these AGN. In the case of log($U$) = $-$2.0, the density profile will decrease as $\propto 1/r^2$ with distance from the nucleus.

This choice of density profile is physically motivated by the observation that the [O~III]/H$\beta$ ratio, which is a strong function of the ionization parameter, is approximately constant across the NLR for most galaxies in the sample (see also \citealp{Stern2014, Wang2022}). An underlying assumption of this method is that the total ionized gas mass is traced out equally well by the [O~III] emission at all radial distances. This type of density profile is generally consistent with the medium ionization Cloudy model component for our galaxies, as shown by the green model points in Figure~\ref{densities}.

\begin{figure}[t!]
\centering
\vspace{1em}
\includegraphics[width=0.47\textwidth]{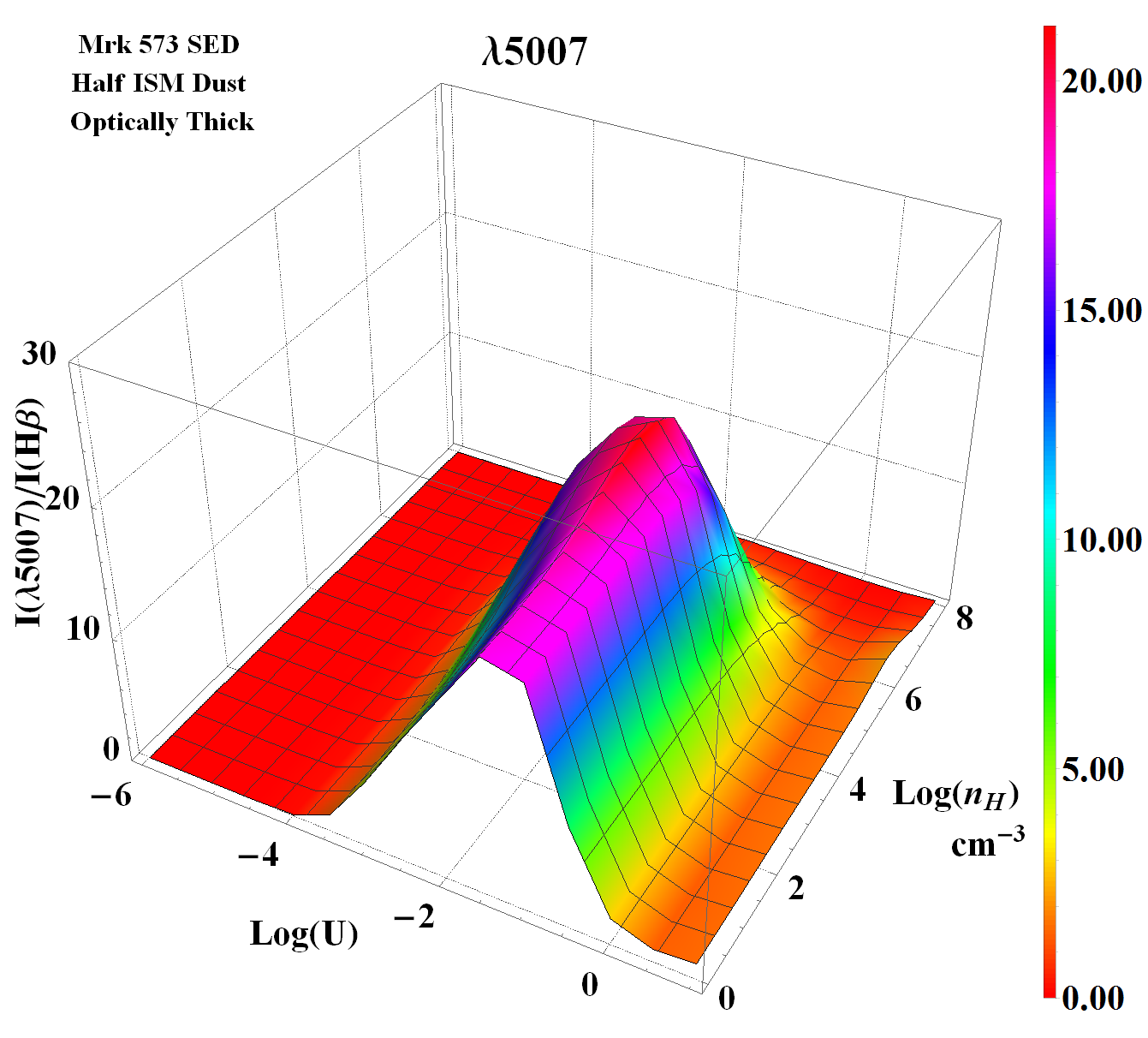}\\\vspace{1.5em}
\includegraphics[width=0.47\textwidth]{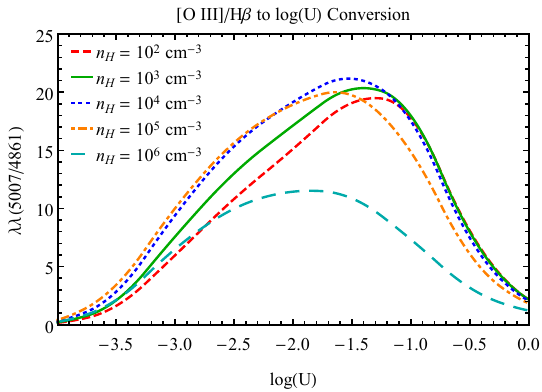}
\caption{The [O~III]/H$\beta$ emission line ratio as a function of ionization parameter ($U$) and density ($n_\mathrm{H}$) for optically-thick gas. The lower panel displays a 2D projection of the upper panel. In the NLR, commonly observed values of [O~III]/H$\beta \approx 5-20$ correspond to changes in $U$ by a factor of $\sim$40 (1.6 dex). Note that it is possible to have the same value of [O~III]/H$\beta$ for two distinct log($U$) values. The data shown in this figure are tabulated in Appendix~\ref{appendix2c}.}
\label{3do3}
\end{figure}

The critical aspect of this method is to choose an appropriate value of the ionization parameter. In this case, we chose various values of the ionization parameter that are known from general Cloudy models to produce the observed [O~III]/H$\beta$ ratios, and calculated the density at each radial distance using the ionization parameter relation (Equation~\ref{uion}). A first approximation is to choose the commonly used value of log($U$) = $-$2.0, which will produce [O~III]/H$\beta$ ratios comparable to those we observe in these galaxies. It is worth noting that this is a degenerate parameter space, with the same value of [O~III]/H$\beta$ possible for multiple ionization parameters if the gas is more or less ionized. This behavior is demonstrated in Figure~\ref{3do3}, which illustrates the reverse-saddle nature of the degeneracy. We therefore choose a range of values around log($U$) = $-$2.0, resulting in the density profiles shown in Figure~\ref{densities}.

\begin{figure*}[ht!]
\vspace{1.0em}
\centering
\includegraphics[scale=0.97]{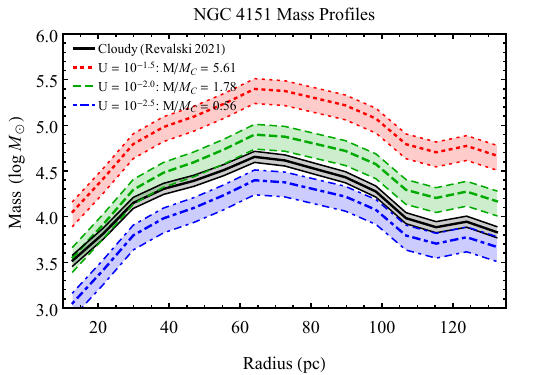}
\includegraphics[scale=0.97]{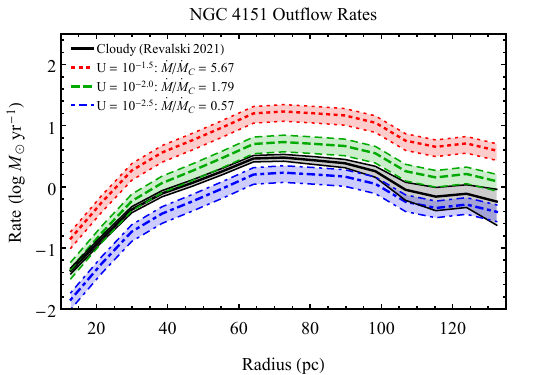}\\\vspace{1.0em}
\includegraphics[scale=0.97]{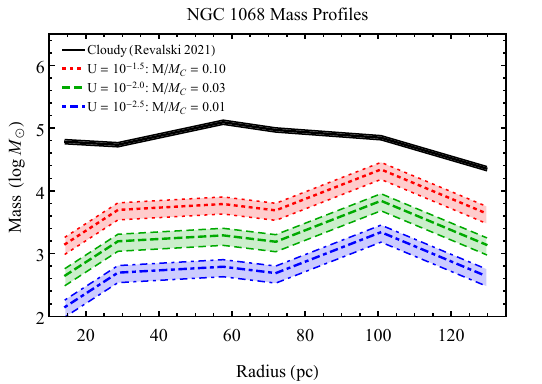}
\includegraphics[scale=0.97]{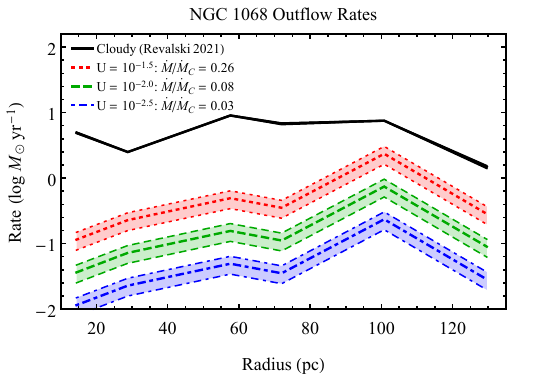}\\\vspace{1.0em}
\includegraphics[scale=0.97]{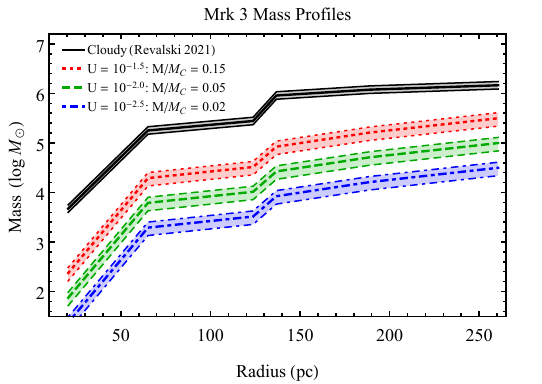}
\includegraphics[scale=0.97]{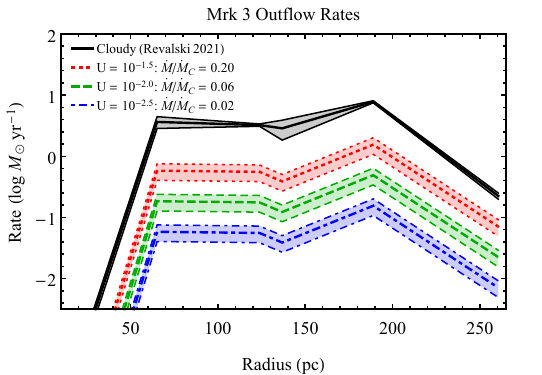}
\caption{The ionized gas mass (left column) and outflow rate (right column) profiles for NGC~4151, NGC~1068, Mrk~3, Mrk~573, Mrk~78, and Mrk~34, calculated from Cloudy models in solid black (\citealp{Revalski2021}), and using the recombination relation (Equation~\ref{msimp}), assuming constant ionization parameters with radius of log($U$) = $-$1.5 (dotted red), log($U$) = $-$2.0 (dashed green), and log($U$) = $-$2.5 (dashed-dotted blue). The legend indicates the linear ratio of the total constant density mass to the Cloudy derived mass ($M/M_C$) and peak outflow rates ($\dot M/ \dot M_C$). The uncertainties in the simplified method are dominated by the uncertainty in $\alpha_{_{H\beta}}^{eff}$ due to its dependence on temperature. Overall, using the recombination relation with a constant $U(r)$ shows significant scatter between objects, due to the different levels of ionization in each NLR.}
\vspace{5.0em}
\label{uionrecomb}
\end{figure*}
\addtocounter{figure}{-1}
\begin{figure*}[ht!]
\vspace{1.0em}
\centering
\includegraphics[scale=0.97]{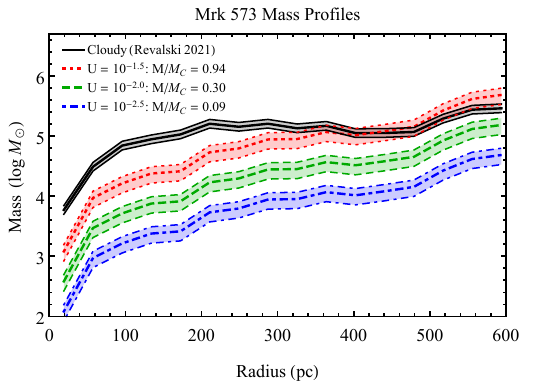}
\includegraphics[scale=0.97]{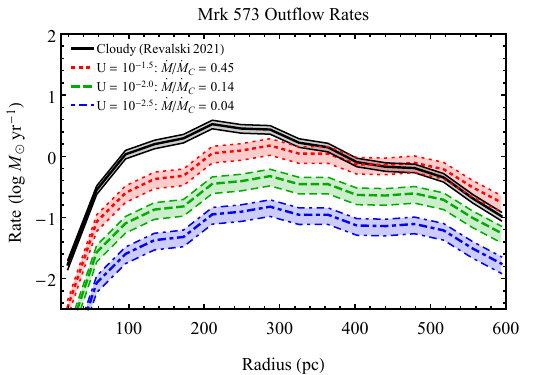}\\\vspace{1.0em}
\includegraphics[scale=0.97]{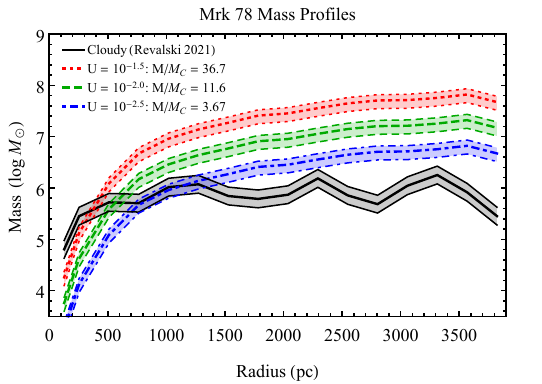}
\includegraphics[scale=0.97]{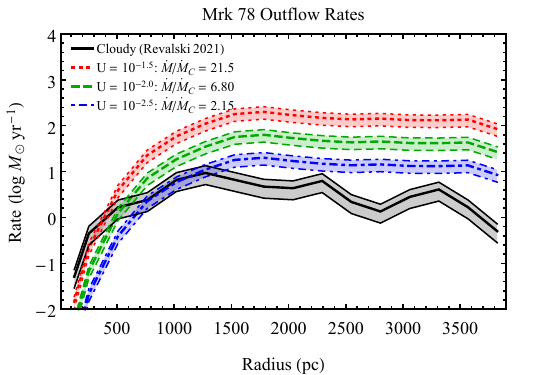}\\\vspace{1.0em}
\includegraphics[scale=0.97]{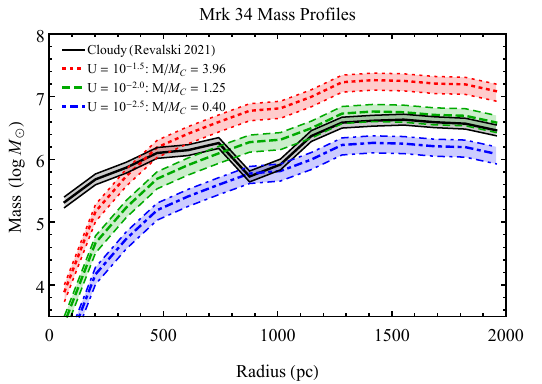}
\includegraphics[scale=0.97]{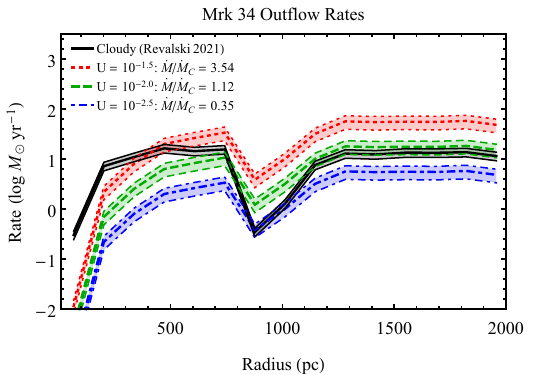}
\caption{{\it continued.}}
\end{figure*}

The results for log($U$) = [$-$1.5, $-$2.0, $-$2.5] are shown in Figure~\ref{uionrecomb}. This procedure produces a reasonable match for NGC~4151, Mrk~573, and Mrk~34, albeit with a higher value for Mrk~573, but disagrees with the benchmark results by more than 1 dex for NGC~1068, Mrk~3, and Mrk~78. To better understand the variations in $U$ that are required to produce an acceptable match, we show several ionization parameters in Figure~\ref{uionrecomb}. The cases where this technique disagrees most severely may be attributed to the fact that the gas is not a fully ionized medium as assumed by radiative recombination, and that the ionization parameter is not constant with radius. In the next sections, we test this by using the same power-law density profiles, as well as profiles that vary in ionization as a function of radius, with single-component, optically-thick photoionization models.

\subsection{Photoionization Models with Constant U(r)}\label{sec:uphotcon}

The fourth method that we consider uses the same power-law density profiles as the previous technique, but these are now input parameters for single-component, optically-thick photoionization models, rather than being used with the recombination relation. The gas masses are then calculated using Equation~\ref{masseqn}. In the cases of Mrk~573, Mrk~78, and Mrk~34, we used the exact photoionization models with log($U$) = [$-$1.5, $-$2.0, $-$2.5] from \cite{Revalski2018a, Revalski2018b, Revalski2021}. However, NGC~4151, NGC~1068, and Mrk~3 were modeled previously by \cite{Crenshaw2015}, \cite{Kraemer2000b}, and \cite{Collins2009}, respectively. In these cases, we generated new models with the Cloudy spectral synthesis code (version 13.04; \citealp{Ferland2013}), using identical SED, abundance, and dust parameters as our previous studies \citep{Revalski2021}, for the ionizing photon luminosities listed in Table~\ref{sample}.

The results for log($U$) = [$-$1.5, $-$2.0, $-$2.5] are shown in Figure~\ref{uioncloudy}. This procedure produces a reasonable match for NGC~4151, NGC~1068, Mrk~3, Mrk~573, and Mrk~34, for particular values of $U$, with considerably less scatter than when using the recombination relation. However, it is not clear in advance (a priori) what value of $U$ should be chosen, and significant disagreement is still seen for Mrk~78. In this analysis, we adopted the mean [O~III]/H$\beta$ ratio for each galaxy, while noting that they change by factors of $\sim$2~$-$~3 across the NLR in some cases. This is most significant for Mrk~78, where the [O~III]/H$\beta$ ratio decreases from $\sim$18 to $\sim$6 with increasing distance from the nucleus. While accounting for these changes would only make a marginal difference in the approximated H$\beta$ fluxes, it can be seen from Figure~\ref{3do3} that moving from an [O~III]/H$\beta$ ratio of $\sim$18 to $\sim$6 corresponds to a decrease in ionization parameter of log($U$)$~\approx-1.9$ to log($U$)$~\approx-3.2$. Thus a decrease in the [O~III]/H$\beta$ ratio by a factor of $\sim$3 is indicative of a decrease in $U$, and thus an increase in the density, by a factor of $\sim$20.

Using these results, it is clear that this process will only work well with an appropriate estimate of $U$ for each galaxy, and possibly also accounting for changes in $U$ as a function of radial distance. In this context, we explore in the next section using the [O~III]/H$\beta$ ratios for each AGN to constrain $U$. This ratio was chosen because the lines are closely separated, and so are insensitive to reddening, and they are present in a wide variety of spectra. The ratio is relatively insensitive to other physical conditions at $n_\mathrm{H} \lesssim$~10$^5$ cm$^{-3}$, which has made it a powerful diagnostic in BPT diagrams that differentiate sources of ionization  \citep{Baldwin1981, Veilleux1987}.

\subsection{Photoionization Models with Variable U(r)}\label{ssec:uphotvar}

The fifth and final case we explore is similar to the previous, with the gas masses calculated from the single-component photoionization models and Equation~\ref{masseqn}. In this case, we chose a best value of the ionization parameter for each AGN based on the [O~III]/H$\beta$ ratios, and allowed it to vary as a function of radius if indicated by changes in the [O~III]/H$\beta$ ratios. Using the values from Figure~\ref{3do3}, which are tabulated in Appendix~\ref{appendix2c}, we converted the [O~III]/H$\beta$ ratios for each NLR to corresponding $U$ values at each radial distance.

In the case of NGC~4151, the ionization parameter decreases monotonically from log($U$)~$\approx-2.2$ to $-$3.0 moving from zero to 135~pc. For NGC~1068, the [O~III]/H$\beta$ ratios show considerable variations from $\sim$5~$-$~20 between different kinematic components, as shown in Figure~1 of \cite{Kraemer2000b}. The blue-shifted component has a significantly higher flux and we adopt the larger [O~III]/H$\beta$ ratios of that component, resulting in a slight increase from log($U$)~$\approx$~$-$2.35 to $-$1.95 from zero to 130~pc. Similarly, the [O~III]/H$\beta$ ratios for Mrk~3 indicate a small decrease of log($U$)~$\approx$~$-$2.0 to $-$2.4 from zero to 260~pc. In the case of Mrk~573 there were insufficient line ratios to create multi-component models at $r >$~175~pc \citep{Revalski2018a}, but the [O~III]/H$\beta$ ratios indicate an approximately constant log($U$)~$\approx$~$-$2.45 across the NLR. The radial change is most significant for Mrk~78, where the ionization parameter decreases from log($U$)$~\approx$~$-$1.9 to $-$3.2 across the radial extent of the outflows. Finally, the ratios for Mrk~34 are consistent with an approximately constant log($U$)~$\approx$~$-$2.65 at all radii.

It is important to note that additional line ratios could be used to more tightly constrain the ionization parameters, and reduce degeneracies with metallicity and dust content, such as the technique developed by \cite{Baron2019b} that uses the [O~III]/H$\beta$ and [N~II]/H$\alpha$ ratios. However, in this case we are interested in exploring techniques that can be implemented with just a few emission lines over a narrow spectral range. Using additional line ratios would require significant spectral coverage such that we could construct full multi-component photoionization models, negating the purpose of these simplified methods, except for the advantage of saving computational and analysis time. For this reason, we limit our exploration of variable ionization parameters to those that can be obtained from the [O~III]/H$\beta$ ratios.

\begin{figure*}[ht!]
\vspace{1.0em}
\centering
\includegraphics[scale=0.97]{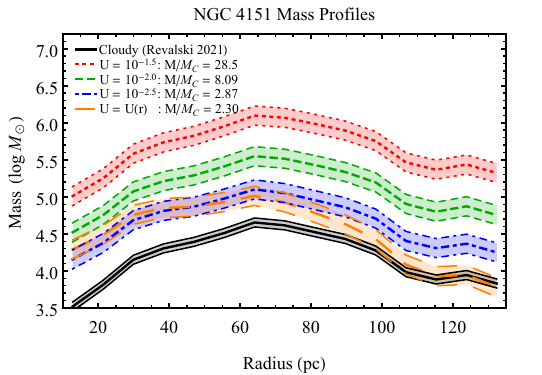}
\includegraphics[scale=0.97]{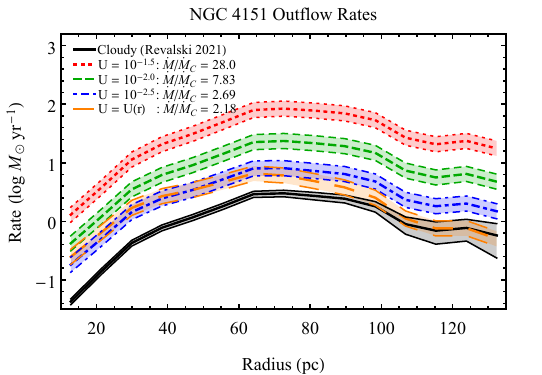}\\\vspace{1.0em}
\includegraphics[scale=0.97]{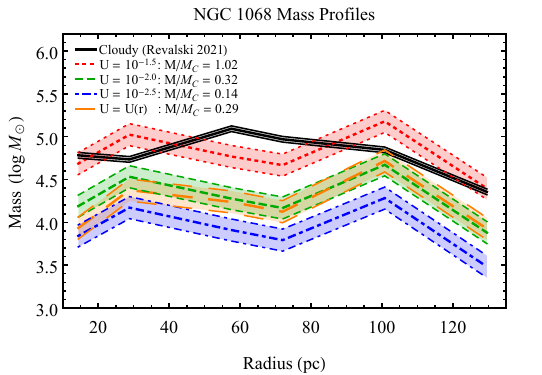}
\includegraphics[scale=0.97]{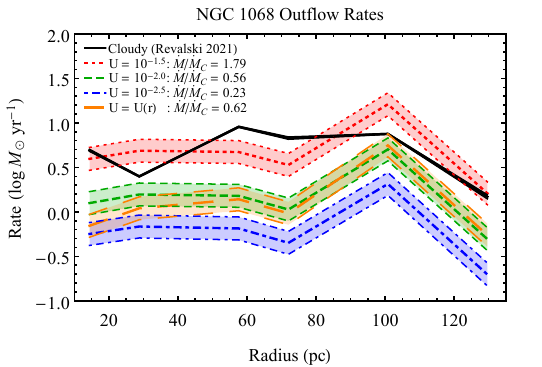}\\\vspace{1.0em}
\includegraphics[scale=0.97]{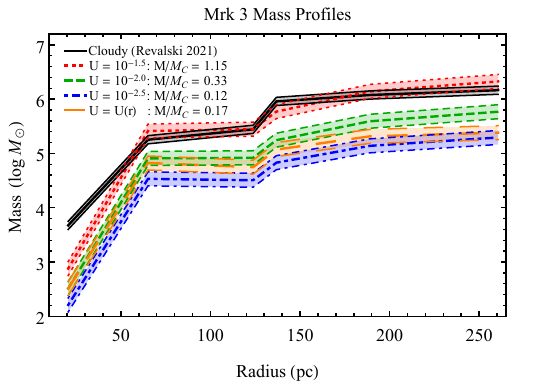}
\includegraphics[scale=0.97]{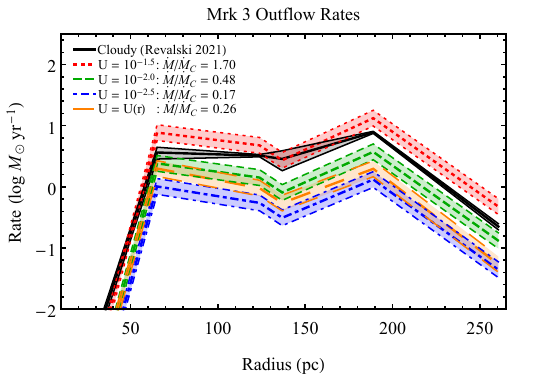}
\caption{The ionized gas mass (left column) and outflow rate (right column) profiles for NGC~4151, NGC~1068, Mrk~3, Mrk~573, Mrk~78, and Mrk~34, calculated from Cloudy models in solid black (\citealp{Revalski2021}), and using single-component photoionization models (Equation~\ref{masseqn}), assuming constant ionization parameters with radius of log($U$) = $-$1.5 (dotted red), log($U$) = $-$2.0 (dashed green), log($U$) = $-$2.5 (dashed-dotted blue), and with log($U$) as a function of radius (long dashed orange, see S\ref{ssec:uphotvar}). The legend indicates the linear ratio of the total single-component mass to the Cloudy derived mass ($M/M_C$) and peak outflow rates ($\dot M/ \dot M_C$). The uncertainties in the simplified method are dominated by the $\pm$0.1~dex uncertainty in converting the [O~III]/H$\beta$ ratios to log($U$) values, and adopting a typical uncertainty of 30\% in the [O~III]/H$\beta$ ratios. It is not clear a priori what log($U$) value should be chosen for a given AGN, as there is no constraint on the column density of the gas. Therefore, using the [O~III]/H$\beta$ ratios to derive log($U$) as a function of radius typically yields the best agreement with the multi-component Cloudy models.}
\vspace{4.0em}
\label{uioncloudy}
\end{figure*}
\addtocounter{figure}{-1}
\begin{figure*}[ht!]
\vspace{0.5em}
\centering
\includegraphics[scale=0.97]{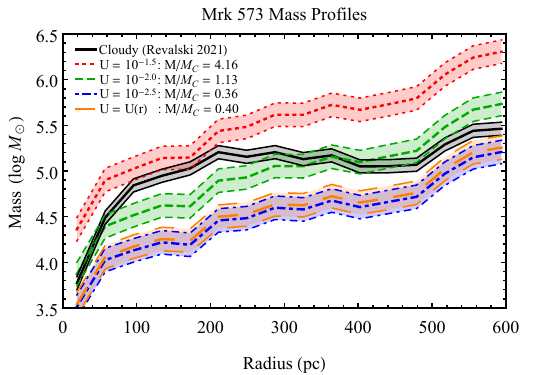}
\includegraphics[scale=0.97]{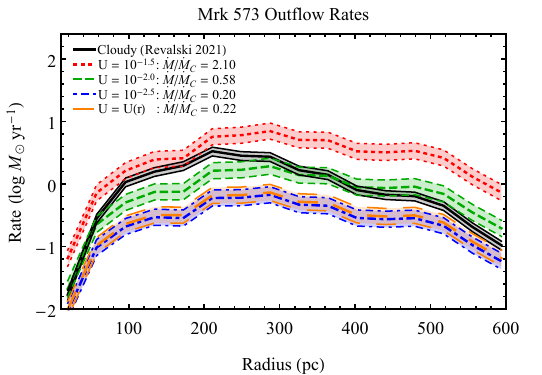}\\\vspace{0.5em}
\includegraphics[scale=0.97]{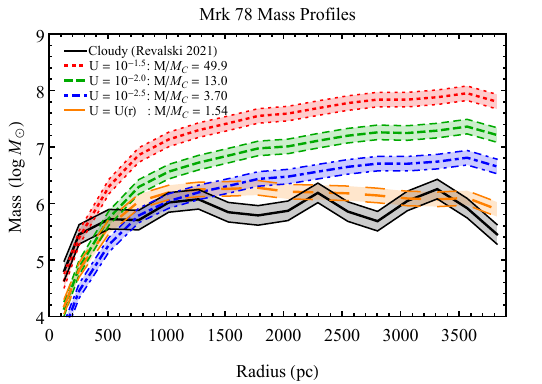}
\includegraphics[scale=0.97]{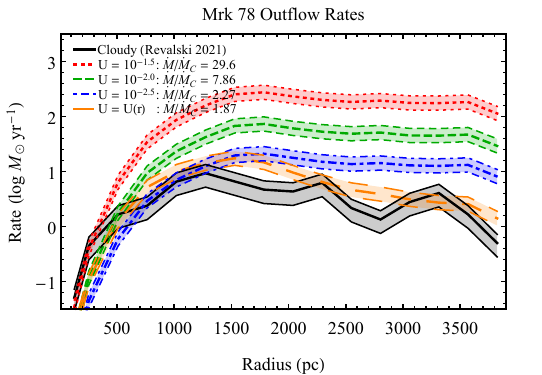}\\\vspace{0.5em}
\includegraphics[scale=0.97]{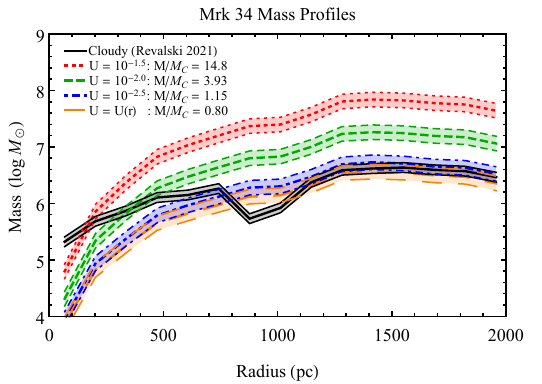}
\includegraphics[scale=0.97]{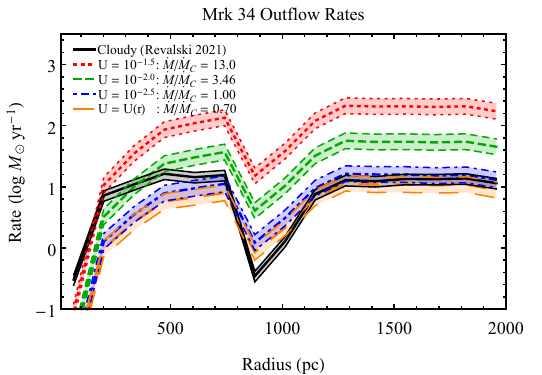}
\vspace{-0.5em}
\caption{{\it continued.}}
%\vspace{1.0em}
\end{figure*}

The results of this process are shown by the long-dashed orange curves in Figure~\ref{uioncloudy}. Overall, this process is able to reproduce the total ionized gas masses and peak outflow rates from our earlier study to within better than a factor of $\sim$4~$-$~5 across the sample. While this may be expected because the process is most similar in nature to multi-component photoionization modeling, it is encouraging that the two processes yield similar results given the vast number of simplifications that are introduced by the single-component models. Specifically, this process assumes that all of the gas is optically-thick, and can be reasonably approximated by a single ionization parameter and density at each location. This also eliminates differences in dust content, gas depletions, and filtered SEDs, that were used in the multi-component analysis.

While these results are encouraging, it is also important to acknowledge the limitations of this process. Specifically, while we are able to excellently reproduce our results for the higher luminosity targets with more extended NLRs, the complex nature in the circumnuclear regions of NGC~1068 and Mrk~3 show residual deviations. In the case of NGC~1068, there are several distinct kinematic components near the nucleus, each of which exhibits a unique [O~III]/H$\beta$ ratio and centroid velocity. In addition, \cite{Kraemer2000b} found evidence of a localized shock, which increases the ionizing flux in small regions. The largest deviation is seen for Mrk~3, where the NLR is being externally fueled as the host galaxy tidally siphons cold gas from a neighboring companion galaxy. This complicates the gas geometry, and leads to a more centrally concentrated and highly ionized NLR \citep{Bogdan2017, Gnilka2020}. In this case, the best fit log($U$)~$\approx$~$-$1.3 resides on the higher ionization slope of the ionization diagram (see Figure~\ref{3do3}), which is indistinguishable from log($U$)~$\approx$~$-$2.5 for [O~III]/H$\beta \approx$~15~$-$~20 without additional line ratios. The agreement is substantially better when using a partially-absorbed SED; however, this also requires additional line ratios to constrain.

This provides perspective on the issue of resolution versus NLR complexity, as the numerous locations modeled along the NLRs of NGC~4151 and NGC~1068 would fit entirely within the central bins extracted for the higher redshift targets, such as Mrk~34 and Mrk~78. This may explain why the simplified techniques are able to better reproduce the multi-component results for the higher luminosity AGN with more extended NLRs, as we are resolving more orderly, large scale gas structures that likely originate from illumination of the host galaxy disk, rather than inner knots of circumnuclear gas.

\section{Discussion}

\begin{deluxetable*}{@{\extracolsep{3.5pt}}lccccccccccccc}
\vspace{0.5em}
\tabletypesize{\footnotesize}
\setlength{\tabcolsep}{0.0in}
\def\arraystretch{1.0}
\tablecaption{Summary of the Simplified Method Results ($M/M_{\mathrm{C}}$ and $\dot M/ \dot M_{\mathrm{C}}$)}
\tablehead{
\colhead{} & \colhead{} & \colhead{} & \multicolumn{7}{c}{\textbf{Recombination}} & \multicolumn{4}{c}{\textbf{Photoionization}}\\[-0.5em]
\cline{4-10} \cline{11-14}
\colhead{Catalog} & \colhead{Total $M_{ion}$} & \colhead{$\dot M_{max}$} & \multicolumn{3}{c}{Constant Density} & \colhead{[S~II]} & \multicolumn{3}{c}{log($U$)} & \multicolumn{4}{c}{log($U$)}\\[-0.5em]
\cline{4-6} \cline{7-7} \cline{8-10} \cline{11-14}
\colhead{Name} & \colhead{(log $M_{\odot}$)} &\colhead{($M_{\odot}$ yr$^{-1}$)} & \colhead{$n_\mathrm{H}=$~10$^2$} & \colhead{$n_\mathrm{H}=$~10$^3$} &\colhead{$n_\mathrm{H}=$~10$^4$} & \colhead{$n_\mathrm{H}(r)$} & \colhead{$-$1.5} & \colhead{$-$2.0} & \colhead{$-$2.5} & \colhead{$-$1.5} & \colhead{$-$2.0} & \colhead{$-$2.5} & \colhead{$U(r)$}\\[-0.5em]
\colhead{(1)} & \colhead{(2)} & \colhead{(3)} &\colhead{(4)} & \colhead{(5)} & \colhead{(6)} & \colhead{(7)} & \colhead{(8)} & \colhead{(9)} & \colhead{(10)} & \colhead{(11)} & \colhead{(12)} & \colhead{(13)} & \colhead{(14)}
}
\startdata
NGC 4151 & 	5.5	$\pm$	0.1	 & 	3.0	$\pm$	0.5 & 57.9/40.0 & 5.79/4.00 & 0.58/0.40 & 6.44/6.18 & 5.61/5.67 & 1.78/1.79 & 0.56/0.57 & 28.5/28.0 & 8.09/7.83 & 2.87/2.69 & 2.30/2.18 \\
NGC 1068 & 	5.6	$\pm$	0.1	 & 	9.0	$\pm$	1.1 & 17.6/25.5 & 1.76/2.55 & 0.18/0.25 & 2.10/3.03 & 0.10/0.26 & 0.03/0.08 & $<$0.01/0.03 & 1.02/1.79 & 0.32/0.56 & 0.14/0.23 & 0.29/0.62 \\
Mrk 3 & 	6.6	$\pm$	0.1	 & 	7.8	$\pm$	1.2 & 2.16/7.30 & 0.22/0.73 & 0.02/0.07 & 1.77/3.67 & 0.15/0.20 & 0.05/0.06 & 0.02/0.02 & 1.15/1.70 & 0.33/0.48 & 0.12/0.17 & 0.17/0.26 \\
Mrk 573 & 	6.3	$\pm$	0.1	 & 	3.4	$\pm$	0.6 & 5.98/5.13 & 0.60/0.51 & 0.06/0.05 & 1.36/1.11 & 0.94/0.45 & 0.30/0.14 & 0.09/0.04 & 4.16/2.10 & 1.13/0.58 & 0.36/0.20 & 0.40/0.22 \\
Mrk 78 & 	7.1	$\pm$	0.1	 & 	9.3	$\pm$	4.6 & 3.03/3.25 & 0.30/0.32 & 0.03/0.03 & 0.76/0.96 & 36.7/21.5 & 11.6/6.80 & 3.67/2.15 & 49.9/29.6 & 13.0/7.86 & 3.70/2.27 & 1.54/1.87 \\
Mrk 34 & 	7.2	$\pm$	0.1	 & 	12.5	$\pm$	2.7 & 2.32/4.44 & 0.23/0.44 & 0.02/0.04 & 0.25/0.42 & 3.96/3.54 & 1.25/1.12 & 0.40/0.35 & 14.8/13.0 & 3.93/3.46 & 1.15/1.00 & 0.80/0.70
\enddata
\tablecomments{A summary of the results presented in Figures~\ref{constrecomb}$-$\ref{uioncloudy}. The first three columns provide the galaxy names, total ionized gas masses, and peak outflow rates from the Cloudy modeling results of \cite{Revalski2021}. The subsequent columns list the ratio of the total simplified technique mass to the Cloudy derived mass ($M/M_{\mathrm{Cloudy}}$), and the ratio of the peak outflow rates ($\dot M/ \dot M_{\mathrm{Cloudy}}$), separated by a slash. The mean, minimum, and maximum of each mass ratio column are shown in Figure~\ref{fig:summary}.}
\label{simptable}
\vspace{-1.0em}
\end{deluxetable*}

\begin{figure}%[hb!]
\vspace{0.5em}
\centering
\includegraphics[width=0.49\textwidth]{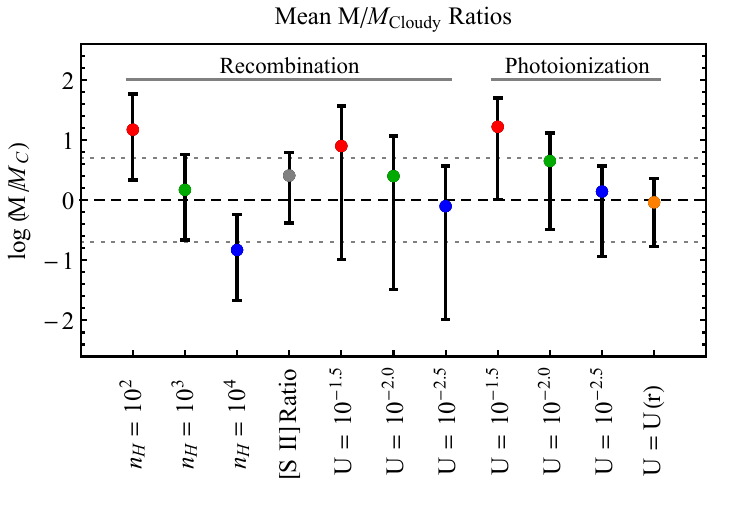}\\
\vspace{-1.5em}
\caption{For each technique, this graphic shows the average $M/M_{\mathrm{Cloudy}}$ ratio for the sample (colored points), along with the minimum and maximum values. The gray dotted lines denote factor-of-five boundaries. In general, points above the dashed unity line indicate that those techniques typically overpredict the gas mass, while points below indicate those techniques typically underpredict the gas mass. Overall, the $n_\mathrm{H} \propto U(r)$ profiles used with single-component photoionization models best reproduce the multi-component results.}
\label{fig:summary}
\end{figure}

We have investigated the effects of deriving ionized gas masses and outflow rates using methodologies that are not based on multi-component photoionization models. The results of these methods differ from our benchmark results primarily due to the techniques used to derive the density laws, and differences in converting the luminosities to masses. These results are summarized in Table~\ref{simptable}, which contains the ratios of the total simplified method masses to Cloudy model masses ($M/M_C$), as well as the ratios of the peak outflow rates ($\dot M / \dot M_C$). In addition, we provide the average, minimum, and maximum gas mass values for the sample, for each technique, in Figure~\ref{fig:summary}. In the case of the outflow rates, the peaks may occur at different radii between the methods, which is important for understanding where energy is being deposited into the host galaxy environment.

Overall, the common assumption of a constant density profile with $n_\mathrm{H} =$~10$^2$ cm$^{-3}$ produces the worst results, overestimating the gas masses and outflow rates in all cases. This trend would also apply to the kinetic energy and momenta flow rates, although the overestimation can be slightly less severe in some cases because it is proportional to $\dot M$ rather than $M$. The conclusion that adopting a density of $n_\mathrm{H} =$~10$^2$ cm$^{-3}$ will overestimate the gas masses has also been noted by others \citep{Perna2017, Baron2019b}, and a mean density of $n_\mathrm{H} \approx$~10$^3$ cm$^{-3}$ produces results that are more consistent with the photoionization models. However, there is a significant range of densities present across each NLR, so adopting a mean gas density is not recommended. These results suggest that claims of very energetic NLR outflows in nearby AGN that are based on global or constant density techniques should be considered with caution, as also suggested by others \citep{Karouzos2016, Bischetti2017, Perna2017}.

The [S~II] density law method is physically motivated as it employs a direct tracer of the gas density. However, these diagnostic lines trace the electron density, which may not adequately represent the hydrogen density in partially ionized zones \citep{Kraemer2000c, Kewley2019, Davies2020Ric, Riffel2021}. Furthermore, due to absorption at smaller radii, and emission arising from outside the primary bicone, we find that these factors lead to [S~II] underestimating the hydrogen density, particularly at small radii. While this technique appears to perform better at larger radii, this is primarily because the NLR densities decrease to the electron density range that is probed by the [S~II] doublet.
 
The assumption of a constant ionization parameter also yields potentially promising results, with the best agreement found using single-component, optically-thick models with $U$ determined from the [O~III]/H$\beta$ ratios. Interestingly, we may have expected the log($U$) = $-$2.0 case to perform better for all galaxies, given the fact that the medium ionization components almost universally follow a $1/r^2$ law, as seen by the green curves in Figure~\ref{densities}. In these cases, the medium ionization component contains between 20\% to 95\% of the total ionized gas mass for each outflow. In this context, the single-component models should not overestimate the mass in any case. However, when running the single-component models, there is no constraint on whether the gas is optically-thin or optically-thick. In the multi-component models, the medium ionization component was generally close to being optically-thick, but not in all cases, especially at small radial distances. Therefore, the dispersion in the best-fitting log($U$) at values below $-$2 is most likely driven by the need to reduce the overall column densities of the single-component models.

For these reasons, determining the value of $U$ at each radius from the [O~III]/H$\beta$ ratios (see Figure~\ref{3do3}) is the most self-consistent technique investigated here that best reproduces the multi-component Cloudy model results. While further refinement of the ionization parameters and column densities are possible with additional emission line diagnostics, this would require sufficient spectral coverage such that running full multi-component photoionization models would be possible, negating the need for a simplified methodology.

The differences between the results of these methods and our benchmark results highlights the need for multi-component photoionization models to fully account for the multiphase nature of the optical emission line gas, when high S/N spectral observations covering a wide range of wavelengths are available. For example, while we are able to reproduce the overall gas masses and peak outflow rates to within factors of a few, and the total mass and energetic correlations shown in Figure~14 of \cite{Revalski2021} would be preserved, the subtle correlation between the bolometric luminosity and mass outflow rates would be heavily skewed and dispersed. This is primarily due to neglecting the relative differences in the ionization states, reddening-corrections, and dust levels at each location in the NLR. We mention in brief that the high spatial resolution HST STIS spectra, which isolate individual bright knots, often yield [O~III]/H$\beta$ ratios that are up to $\sim$40\% larger in some cases, as compared to ground-based observations with substantially larger apertures.

Finally, we consider the implications of our results for calculating global outflow rates, which approximate the mass, velocity, and radial extent of the outflow with a single value for each quantity across the full spatial extent of the outflow. These estimates are subject to strong selection biases, as the emission line centroid, luminosity, and velocity dispersion are weighted towards the optimal conditions for emission; namely, high density or large area regions. In examining the density profiles in Figure~\ref{densities}, and the [O~III] fluxes from our previous studies \citep{Revalski2018a, Revalski2018b, Revalski2019b, Revalski2021}, a significant fraction of the total luminosity is contributed by the dense central knots of emission. These regions trace a minuscule amount of the gas mass, as they emit very efficiently, proportional to the density squared. However, the spatially-integrated spectra of the NLR are often weighted towards the larger area of emission at large radii, such that adopting a lower density that is appropriate for large radii will significantly overestimate the mass of the outflows at small radii. This can be partially compensated for by adopting a higher value for the density ($n_\mathrm{H}\approx$~10$^3$ cm$^{-3}$; \citealp{Baron2019b, Davies2020Ric, Kakkad2020}), and is worthy of further investigation.\\

\section{Conclusions}

We present the first comparison of spatially resolved gas masses and outflow rates derived from multi-component photoionization models \citep{Revalski2021} with techniques commonly used in the literature that have less stringent data requirements. We used a recombination equation with constant densities, those derived from [S~II], and assuming constant ionization parameters. We also explored using single-component photoionization models with constant and variable ionization parameters at each radius. Our conclusions are the following:

\begin{enumerate}
\item In general, multi-component photoionization models provide the most well-constrained measurements of ionized gas masses and outflow rates by accounting for ionization and density stratification at each radial distance along the outflows. However, this technique requires high S/N, spatially-resolved spectroscopy across a wide range of wavelengths, and time-intensive modeling.

\item Using a constant density of $n_\mathrm{H} =$~10$^2$~cm$^{-3}$ overestimates the gas masses and outflow rates for the entire sample. While a higher value of $n_\mathrm{H} =$~10$^3$~cm$^{-3}$ provides better overall agreement to within $\pm$1~dex, there are large variations between objects, and as a function of radius, within the individual outflows.

\item Adopting a constant density with a recombination relation ($M_{ion} \propto L/n_\mathrm{H}$) introduces an artificial correlation with luminosity, which is only physical if the outflow density is independent of radial distance from the SMBH, as well as the AGN luminosity. For these reasons, adopting a constant density is not recommended.

\item Estimating densities from the [S~II] doublet is problematic, because it traces the electron density in a partially-ionized zone that may not trace the [O~III] and H$\beta$ emission. Despite this, use of the [S~II] densities is able to reproduce the total gas masses and peak outflow rates to within $\pm$0.8~dex in most cases, with larger deviations at small radii where the outflow rates often peak.

\item Using the recombination relation with a power-law density profile based on constant ionization parameters is able to reproduce the gas masses well in some cases, but disagrees when there are significant contributions from both optically-thick (radiation bounded) and optically-thin (matter bounded) components, as well as higher ionization gas.

\item Global outflow rates that approximate the mass, velocity, and radial extent of an outflow with single values measured from spatially-integrated spectra are subject to strong selection biases. Primarily, the emission line centroid, dispersion, and luminosity, are weighted towards the dominant emission sampled at large radii. Adopting low densities appropriate for those regions will significantly overestimate the mass and energetics of the outflows in the dense and bright inner regions.

\item The medium ionization component that produces the majority of the [O~III] emission is well-traced by models with log($U$) = $-$2.0. This component generally contains a significant fraction of the gas mass; however, without further line diagnostics to constrain the column density, it is unclear what value of log($U$) should be adopted for each AGN when using single-component models. Thus, using a constant value of log($U$) at all radii only marginally reproduces the gas masses and outflow rates of the multi-component models.

\item We find that single-component, optically-thick photoionization models where log(U) varies as a function of radius based on the [O~III]/H$\beta$ ratios are generally a reliable means of determining the ionized gas masses and outflow rates in systems with insufficient data to construct multi-component models. This requires accepting uncertainties in the gas masses and outflow rates by factors of $\sim$3~($\pm$0.5~dex) in most cases, and up to $\sim$5~($\pm$0.7~dex) at some radii, which may be acceptable for a variety of science goals.\\
\end{enumerate}

\acknowledgements

The authors thank the anonymous referee for helpful comments that improved the clarity of this paper. This work was supported by the STScI Directors Discretionary Research Fund (DDRF) Proposal 82490. This research is based on observations made with the NASA/ESA Hubble Space Telescope obtained from the Space Telescope Science Institute, which is operated by the Association of Universities for Research in Astronomy, Inc., under NASA contract NAS 5–26555. These observations are associated with programs 5140, 5754, 7404, 7573, and 8480. Based in part on observations obtained with the Apache Point Observatory 3.5-meter telescope, which is owned and operated by the Astrophysical Research Consortium.

This paper used the photoionization code Cloudy, which can be obtained from \url{http://www.nublado.org} and the Atomic Line List available at \url{http://www.pa.uky.edu/~peter/atomic/}. This research has made use of the NASA/IPAC Extragalactic Database (NED), which is operated by the Jet Propulsion Laboratory, California Institute of Technology, under contract with the National Aeronautics and Space Administration. This research has made use of NASA's Astrophysics Data System.

\facilities{HST(STIS), ARC(DIS)}

\software{MultiNest \citep{Feroz2019}, 
Cloudy \citep{Ferland2013}, 
Mathematica \citep{Mathematica}, 
Python (\citealp{VanRossum2009}, \url{https://www.python.org}), 
Interactive Data Language (IDL, \url{https://www.harrisgeospatial.com/Software-Technology/IDL}).
}

\bibliography{references}{}
\bibliographystyle{aasjournal}

\appendix
\restartappendixnumbering

\section{Recombination Equation Derivation}
\label{appendix1}

The most common luminosity-based technique used in the literature estimates the gas mass based on the luminosity of a single emission line at constant density. A recombination line such as H$\alpha$ or H$\beta$ is typically chosen, as these are more stable than the forbidden lines across a wide range of physical conditions. When accessible, H$\beta$ is the preferred option as it is less sensitive to collisional effects than H$\alpha$, and is generally not blended with other strong emission lines. This is important, as this technique is essentially ``photon counting'' to determine the mass, which assumes that the emission is dominated by pure radiative recombination and neglects other emission processes. This simplifies the analysis and allows for a single multiplicative factor known as the recombination coefficient to relate the number of photons to the number of hydrogen atoms, and thus the ionized gas mass, i.e. $M \propto L/n_e$.

The main differences between photoionization modeling and this single emission line approach are illuminated by deriving the exact expression that relates the H$\beta$ luminosity to the gas mass. A simplified form of this derivation for pure hydrogen can be found in \cite{Peterson1997}, with additional expressions and physical insight gathered from \cite{Osterbrock2006}. The physical setup is as follows. First, we consider an unresolved region containing discrete gas clouds that each contribute line emission to the observed spectrum. The total ionized mass ($M_{ion}$) in these clouds is given by
\begin{equation}\label{mcloud}
M_{ion} = \frac{4\pi}{3}~l^3~n_e~m_p~N_c
\end{equation}
where $l$ is the radius of a cloud, $n_e$ is the electron density, $m_p$ is the mass of a proton, and $N_c$ is the total number of clouds. This framework establishes the mass by defining the cloud volume ($4\pi l^3/3$) multiplied by the density to get the total number of particles. This result is multiplied by the mass per particle to get the mass of a single cloud, which is then summed over the total number of clouds. While this process uses spherical clouds to define the volume, adaptations for the geometry are also possible.

Next, the emission released by these gas clouds is derived from the gas emissivity ($j_{_{H\beta}}$), which is the luminosity per unit volume per solid angle, defined as
\begin{equation}\label{emiss}
j_{_{H\beta}} = \frac{1}{4\pi}~n_en_p~\alpha_{_{H\beta}}^{eff}~h\nu_{_{H\beta}}~\mathrm{(erg~s^{-1}~cm^{-3}~ster^{-1})}
\end{equation}
where $n_e$ and $n_p$ are the electron and proton number densities, and $\alpha_{_{H\beta}}^{eff}$ is the effective recombination coefficient that describes all transitions from levels $n \geq 4$ that will eventually transition to $n = 2$ and release an H$\beta$ photon. The recombination coefficient is a weak function of temperature due to collisional ionization effects\footnote{The effective recombination coefficient is also a very weak function of density owing to collisional effects, which is increasingly negligible for higher temperatures. Over the density range of $n_\mathrm{H} =$~10$^2 - $10$^6$ cm$^{-3}$ the change is $\sim$4.1\% at $T =$~5,000~K, $\sim$1.7\% at $T =$~10,000~K, and $\sim$0.6\% at $T =$~20,000~K. See Table 4.4 in \cite{Osterbrock2006}.}, approximately following $\alpha_{_{H\beta}}^{eff} \propto T^{-0.9}$. Exact values for various temperatures can be found in Tables 4.2 and 4.4 of \cite{Osterbrock2006}. Assuming optically thick (Case B) recombination:
\begin{eqnarray}
\alpha_{_{H\beta}}^{eff} &=& 5.37\times10^{-14}~\mathrm{(cm^3~s^{-1},~T = ~5,000~K)}\nonumber\\\nonumber
&=& 3.03\times10^{-14}~\mathrm{(cm^3~s^{-1},~T = 10,000~K)}\\\nonumber
&=& 2.10\times10^{-14}~\mathrm{(cm^3~s^{-1},~T = 15,000~K)}\\\nonumber
&=& 1.62\times10^{-14}~\mathrm{(cm^3~s^{-1},~T = 20,000~K)}\nonumber
\end{eqnarray}
The emitted luminosity is then the emissivity integrated over all angles ($d\Omega$) and volume ($dV$),
\begin{eqnarray}
L_{\mathrm{H}\beta} &=& \int\int j_{_{H\beta}}~d\Omega~dV\\ &=&\nonumber \frac{1}{4\pi}~n_en_p~\alpha_{_{H\beta}}^{eff}~h\nu_{_{H\beta}} \times 4\pi \times \frac{4\pi}{3}~N_c~l^3\\ &=&\nonumber \frac{4\pi}{3}~N_c~l^3~n_en_p~\alpha_{_{H\beta}}^{eff}~h\nu_{_{H\beta}}.
\end{eqnarray}
Using the original expression for the total gas mass (Equation~\ref{mcloud}) we can identify the first portion of this expression as $M/m_p$ and write the luminosity as
\begin{equation}
L_{\mathrm{H}\beta} = \frac{4\pi}{3m_p}Mn_p~\alpha_{_{H\beta}}^{eff}~h\nu_{_{H\beta}}.
\end{equation}
Solving for the mass and introducing $m_p = m_p^{eff}$ and $n_p = n_p^{eff}$ to account for elements other than hydrogen,
\begin{equation}\label{mtheory}
M_{ion} = \left(\frac{L_{\mathrm{H}\beta}}{\alpha_{_{H\beta}}^{eff}~h\nu_{_{H\beta}}}\right)\left(\frac{m_p^{eff}}{n_p^{eff}}\right).
\end{equation}
We found that the gaseous abundances of the NLR for our AGN are $Z_{NLR} \approx 1.3~Z_\odot$ (see also \citealp{Dors2021}), and taking into account elements heavier than hydrogen yields $m_p^{eff} = \mu m_p \approx 1.4 m_p$ and $n_p^{eff} \approx 1.1 n_e$. The remaining physical quantity is the energy of an H$\beta$ photon, which is
\begin{eqnarray}
E &=& h\nu_{_{H\beta}}\nonumber\\ &=& (6.626\times10^{-27}~\mathrm{erg~s}) \times (6.165\times10^{14}~\mathrm{Hz})\nonumber\\ &=& 4.085\times10^{-12}~\mathrm{erg} \approx 2.55~\mathrm{eV}.\nonumber
\end{eqnarray}
Incorporating these constants into Equation~\ref{mtheory} yields the expression relating the H$\beta$ luminosity to the ionized gas mass
\begin{equation}
M_{ion} = 5.21 \times 10^{-13} \left(\frac{L_{\mathrm{H}\beta}}{\alpha_{_{H\beta}}^{eff} n_e}\right)~\mathrm{(g)},
\end{equation}
or equivalently,
\begin{equation}
M_{ion} = 2.62 \times 10^{-46} \left(\frac{L_{\mathrm{H}\beta}}{\alpha_{_{H\beta}}^{eff} n_e}\right)~\mathrm{(M_\odot)}.
\end{equation}

We assign uncertainties to these expressions by adopting a range of effective recombination coefficients that are appropriate for the range of temperatures observed in the NLR. Various studies have derived the electron temperature for Seyfert galaxies using the [O~III] $\lambda \lambda$4363/5007 emission line ratio that is sensitive to the gas electron temperature as shown in Figure~\ref{tempdengrid}. Reported NLR temperatures span $T \approx$~5,000~$-$~50,000~K \citep{Dors2015}. However, most studies find a mean [O~III] temperature of $T \approx$~15,000~K with a standard deviation $\sim$2,500~$-$~5,000~K \citep{Bennert2006, Vaona2012, Zhang2013}. We conservatively adopt the upper range as the formal uncertainty, yielding a mean NLR temperature of $T \approx$~15,000~$\pm$~5,000~K.

As discussed in \cite{Revalski2018a}, these temperatures probe the [O~III] emission line gas, and our photoionization models showed that the [S~II] gas is on average $\sim$60\% cooler than the [O~III] gas, which is in excellent agreement with the observational results of \cite{Vaona2012}. If we adopt this lower temperature, then the recombination coefficients increase and the mass estimates decrease by a similar factor of $\sim$1.6. Inserting the effective recombination coefficients that correspond to $T \approx$~15,000~$\pm$~5,000~K, we obtain the useful expressions
\begin{equation}
M_{ion} = (24.81~_{-7.62}^{+7.35})\left(\frac{L_{\mathrm{H}\beta}}{n_e}\right)~\mathrm{(g)},
\end{equation}
and,
\begin{equation}
M_{ion} = (12.48~_{-3.83}^{+3.70}\times10^{-33})\left(\frac{L_{\mathrm{H}\beta}}{n_e}\right)~\mathrm{(M_\odot)}.
\end{equation}

This result is appealing, as it only requires a measurement of the emission line luminosity and the electron density to determine the gas mass within any particular region. However, there are still several observational considerations. Specifically, deriving the intrinsic H$\beta$ luminosity from the observed flux requires an accurate correction for extinction from dust and geometric dilution. In general, at least two hydrogen or helium recombination lines are required to derive the reddening (or color excess, i.e. the differential extinction between the photometric B and V bands due to dust), while correcting for geometric dilution requires an accurate estimate of the distance to the galaxy. In addition, adopting a single electron density for all of the emitting material may not be realistic. Furthermore, the uncertainties in the above equations are entirely due to the dependence of $\alpha_{_{H\beta}}^{eff}$ on a broad range of temperatures ($T \approx$~10,000~$-$~20,000~K) if the latter is not well constrained by emission line diagnostics. Finally, we note that additional refinements are possible, such as accounting for the effects of dust in the gas, which \cite{Baron2019b} found lowers the recombination coefficient by a factor of $\sim$2.

\section{Tabulated Data for Select Figures}

\subsection{Data for the Density Profiles in Figure~\ref{densities}}
\label{appendix2a}
This table provides the values displayed in Figure~\ref{densities}, which are useful for comparing different density estimate techniques.
\startlongtable
\begin{deluxetable*}{@{\extracolsep{3pt}}ccccccccc}
\tabletypesize{\small}
\def\arraystretch{0.96}
\tablecaption{Tabulated Data for the Density Profiles in Figure~\ref{densities}}
\tablehead{
\colhead{Position} & \colhead{Cloudy} & \colhead{Cloudy} & \colhead{Cloudy} & \colhead{[S~II]} & \colhead{log($U$)} & \colhead{log($U$)} & \colhead{log($U$)} & \colhead{log($U(r)$)}\\[-0.5em]
\colhead{(\arcsec)} & \colhead{High} & \colhead{Med} & \colhead{Low} & \colhead{Ratio} & \colhead{= $-$1.5} & \colhead{= $-$2.0} & \colhead{= $-$2.5} & \colhead{n$_\mathrm{H}$~(log($U$))}\\[-0.5em]
\colhead{(1)} & \colhead{(2)} & \colhead{(3)} &\colhead{(4)} & \colhead{(5)} & \colhead{(6)} & \colhead{(7)} & \colhead{(8)} & \colhead{(9)}
}
\startdata
NGC 4151 \\
\hline
-2.47	 & 	\nodata	 & 	\nodata	 & 	\nodata	 & 	\nodata	 & 	1.98	 & 	2.48	 & 	2.98	 & 	\nodata	(\nodata) \\
-2.35	 & 	\nodata	 & 	\nodata	 & 	\nodata	 & 	\nodata	 & 	2.03	 & 	2.53	 & 	3.03	 & 	\nodata	(\nodata) \\
-2.21	 & 	\nodata	 & 	\nodata	 & 	\nodata	 & 	\nodata	 & 	2.07	 & 	2.57	 & 	3.07	 & 	\nodata	(\nodata) \\
-2.09	 & 	1.18	 & 	2.88	 & 	\nodata	 & 	2.30	 & 	2.12	 & 	2.62	 & 	3.12	 & 	3.68	(-3.00) \\
-1.96	 & 	\nodata	 & 	\nodata	 & 	\nodata	 & 	\nodata	 & 	2.18	 & 	2.68	 & 	3.18	 & 	3.68	(-2.95) \\
-1.84	 & 	\nodata	 & 	\nodata	 & 	\nodata	 & 	\nodata	 & 	2.24	 & 	2.74	 & 	3.24	 & 	3.70	(-2.90) \\
-1.71	 & 	1.30	 & 	3.00	 & 	\nodata	 & 	2.60	 & 	2.30	 & 	2.80	 & 	3.30	 & 	3.71	(-2.85) \\
-1.58	 & 	\nodata	 & 	\nodata	 & 	\nodata	 & 	\nodata	 & 	2.37	 & 	2.87	 & 	3.37	 & 	3.73	(-2.80) \\
-1.46	 & 	\nodata	 & 	\nodata	 & 	\nodata	 & 	\nodata	 & 	2.44	 & 	2.94	 & 	3.44	 & 	3.77	(-2.75) \\
-1.33	 & 	1.55	 & 	3.08	 & 	\nodata	 & 	3.12	 & 	2.52	 & 	3.02	 & 	3.52	 & 	3.80	(-2.70) \\
-1.20	 & 	\nodata	 & 	\nodata	 & 	\nodata	 & 	\nodata	 & 	2.60	 & 	3.10	 & 	3.60	 & 	3.80	(-2.70) \\
-1.08	 & 	\nodata	 & 	\nodata	 & 	\nodata	 & 	\nodata	 & 	2.70	 & 	3.20	 & 	3.70	 & 	3.85	(-2.65) \\
-0.95	 & 	1.78	 & 	3.30	 & 	\nodata	 & 	2.95	 & 	2.81	 & 	3.31	 & 	3.81	 & 	3.91	(-2.60) \\
-0.82	 & 	2.00	 & 	3.48	 & 	\nodata	 & 	2.89	 & 	2.93	 & 	3.43	 & 	3.93	 & 	3.98	(-2.55) \\
-0.70	 & 	\nodata	 & 	\nodata	 & 	\nodata	 & 	\nodata	 & 	3.08	 & 	3.58	 & 	4.08	 & 	4.08	(-2.50) \\
-0.57	 & 	2.18	 & 	3.70	 & 	\nodata	 & 	2.80	 & 	3.25	 & 	3.75	 & 	4.25	 & 	4.20	(-2.45) \\
-0.44	 & 	2.57	 & 	4.08	 & 	\nodata	 & 	3.41	 & 	3.47	 & 	3.97	 & 	4.47	 & 	4.37	(-2.40) \\
-0.32	 & 	\nodata	 & 	\nodata	 & 	\nodata	 & 	\nodata	 & 	3.76	 & 	4.26	 & 	4.76	 & 	4.62	(-2.35) \\
-0.19	 & 	2.82	 & 	4.26	 & 	\nodata	 & 	3.19	 & 	4.21	 & 	4.71	 & 	5.21	 & 	5.01	(-2.30) \\
0.19	 & 	4.08	 & 	5.00	 & 	7.00	 & 	3.12	 & 	4.21	 & 	4.71	 & 	5.21	 & 	5.01	(-2.30) \\
0.32	 & 	\nodata	 & 	\nodata	 & 	\nodata	 & 	\nodata	 & 	3.76	 & 	4.26	 & 	4.76	 & 	4.62	(-2.35) \\
0.44	 & 	2.78	 & 	4.08	 & 	\nodata	 & 	3.07	 & 	3.47	 & 	3.97	 & 	4.47	 & 	4.37	(-2.40) \\
0.57	 & 	2.18	 & 	3.70	 & 	\nodata	 & 	2.70	 & 	3.25	 & 	3.75	 & 	4.25	 & 	4.20	(-2.45) \\
0.70	 & 	\nodata	 & 	\nodata	 & 	\nodata	 & 	\nodata	 & 	3.08	 & 	3.58	 & 	4.08	 & 	4.08	(-2.50) \\
0.82	 & 	2.00	 & 	3.48	 & 	\nodata	 & 	2.49	 & 	2.93	 & 	3.43	 & 	3.93	 & 	3.98	(-2.55) \\
0.95	 & 	1.78	 & 	3.30	 & 	\nodata	 & 	3.23	 & 	2.81	 & 	3.31	 & 	3.81	 & 	3.91	(-2.60) \\
1.08	 & 	\nodata	 & 	\nodata	 & 	\nodata	 & 	\nodata	 & 	2.70	 & 	3.20	 & 	3.70	 & 	3.85	(-2.65) \\
1.20	 & 	\nodata	 & 	\nodata	 & 	\nodata	 & 	\nodata	 & 	2.60	 & 	3.10	 & 	3.60	 & 	3.80	(-2.70) \\
1.33	 & 	\nodata	 & 	\nodata	 & 	\nodata	 & 	\nodata	 & 	2.52	 & 	3.02	 & 	3.52	 & 	3.80	(-2.70) \\
1.46	 & 	\nodata	 & 	\nodata	 & 	\nodata	 & 	\nodata	 & 	2.44	 & 	2.94	 & 	3.44	 & 	3.77	(-2.75) \\
1.58	 & 	\nodata	 & 	\nodata	 & 	\nodata	 & 	\nodata	 & 	2.37	 & 	2.87	 & 	3.37	 & 	3.73	(-2.80) \\
1.71	 & 	1.30	 & 	3.12	 & 	\nodata	 & 	2.84	 & 	2.30	 & 	2.80	 & 	3.30	 & 	3.71	(-2.85) \\
1.84	 & 	\nodata	 & 	\nodata	 & 	\nodata	 & 	\nodata	 & 	2.24	 & 	2.74	 & 	3.24	 & 	3.70	(-2.90) \\
1.96	 & 	\nodata	 & 	\nodata	 & 	\nodata	 & 	\nodata	 & 	2.18	 & 	2.68	 & 	3.18	 & 	3.68	(-2.95) \\
2.09	 & 	1.30	 & 	2.88	 & 	\nodata	 & 	2.13	 & 	2.12	 & 	2.62	 & 	3.12	 & 	3.68	(-3.00) \\
2.21	 & 	\nodata	 & 	\nodata	 & 	\nodata	 & 	\nodata	 & 	2.07	 & 	2.57	 & 	3.07	 & 	\nodata	(\nodata) \\
2.35	 & 	\nodata	 & 	\nodata	 & 	\nodata	 & 	\nodata	 & 	2.03	 & 	2.53	 & 	3.03	 & 	\nodata	(\nodata) \\
2.47	 & 	1.18	 & 	2.73	 & 	\nodata	 & 	2.09	 & 	1.98	 & 	2.48	 & 	2.98	 & 	\nodata	(\nodata) \\
\hline
NGC 1068 \\
\hline
-1.67	 & 	3.28	 & 	4.48	 & 	\nodata	 & 	2.25	 & 	3.32	 & 	3.82	 & 	4.32	 & 	3.38	(-1.80) \\
-1.30	 & 	3.28	 & 	4.54	 & 	\nodata	 & 	3.54	 & 	3.54	 & 	4.04	 & 	4.54	 & 	3.49	(-1.80) \\
-0.93	 & 	3.40	 & 	4.61	 & 	\nodata	 & 	2.62	 & 	3.83	 & 	4.33	 & 	4.83	 & 	3.94	(-1.95) \\
-0.74	 & 	\nodata	 & 	\nodata	 & 	\nodata	 & 	2.80	 & 	4.03	 & 	4.53	 & 	5.03	 & 	4.23	(-2.05) \\
-0.37	 & 	4.08	 & 	5.24	 & 	\nodata	 & 	2.91	 & 	4.63	 & 	5.13	 & 	5.63	 & 	4.98	(-2.20) \\
-0.18	 & 	4.08	 & 	5.24	 & 	\nodata	 & 	3.26	 & 	5.23	 & 	5.73	 & 	6.23	 & 	5.73	(-2.35) \\
0.18	 & 	\nodata	 & 	\nodata	 & 	\nodata	 & 	2.83	 & 	5.23	 & 	5.73	 & 	6.23	 & 	5.73	(-2.35) \\
0.37	 & 	3.18	 & 	3.88	 & 	\nodata	 & 	2.43	 & 	4.63	 & 	5.13	 & 	5.63	 & 	4.98	(-2.20) \\
0.74	 & 	4.06	 & 	5.06	 & 	\nodata	 & 	\nodata	 & 	4.03	 & 	4.53	 & 	5.03	 & 	4.23	(-2.05) \\
0.93	 & 	3.98	 & 	4.98	 & 	\nodata	 & 	\nodata	 & 	3.83	 & 	4.33	 & 	4.83	 & 	3.94	(-1.95) \\
1.30	 & 	3.61	 & 	4.61	 & 	\nodata	 & 	3.12	 & 	3.54	 & 	4.04	 & 	4.54	 & 	3.78	(-1.95) \\
1.67	 & 	\nodata	 & 	\nodata	 & 	\nodata	 & 	2.69	 & 	3.32	 & 	3.82	 & 	4.32	 & 	3.38	(-1.80) \\
\hline
Mrk 3 \\
\hline
-0.95	 & 	2.08	 & 	3.50	 & 	2.00	 & 	\nodata	 & 	2.72	 & 	3.22	 & 	3.72	 & 	3.47	(-2.40) \\
-0.69	 & 	2.57	 & 	3.57	 & 	2.25	 & 	2.67	 & 	2.99	 & 	3.49	 & 	3.99	 & 	\nodata	(\nodata) \\
-0.50	 & 	3.16	 & 	4.41	 & 	2.68	 & 	2.17	 & 	3.27	 & 	3.77	 & 	4.27	 & 	3.87	(-2.20) \\
-0.45	 & 	3.16	 & 	4.41	 & 	2.68	 & 	2.17	 & 	3.36	 & 	3.86	 & 	4.36	 & 	3.87	(-2.20) \\
-0.24	 & 	3.67	 & 	4.67	 & 	3.29	 & 	2.14	 & 	3.92	 & 	4.42	 & 	4.92	 & 	4.21	(-2.10) \\
-0.08	 & 	4.93	 & 	4.93	 & 	4.13	 & 	2.65	 & 	4.92	 & 	5.42	 & 	5.92	 & 	7.07	(-2.00) \\
0.08	 & 	4.93	 & 	4.93	 & 	4.13	 & 	2.73	 & 	4.92	 & 	5.42	 & 	5.92	 & 	7.07	(-2.00) \\
0.24	 & 	3.74	 & 	5.24	 & 	3.08	 & 	2.32	 & 	3.92	 & 	4.42	 & 	4.92	 & 	4.21	(-2.10) \\
0.45	 & 	2.74	 & 	4.68	 & 	2.84	 & 	2.12	 & 	3.36	 & 	3.86	 & 	4.36	 & 	3.87	(-2.20) \\
0.50	 & 	2.74	 & 	4.68	 & 	2.84	 & 	2.12	 & 	3.27	 & 	3.77	 & 	4.27	 & 	3.87	(-2.20) \\
0.69	 & 	\nodata	 & 	\nodata	 & 	\nodata	 & 	2.67	 & 	2.99	 & 	3.49	 & 	3.99	 & 	\nodata	(\nodata) \\
0.95	 & 	\nodata	 & 	\nodata	 & 	\nodata	 & 	1.63	 & 	2.72	 & 	3.22	 & 	3.72	 & 	3.47	(-2.40) \\
\hline
Mrk 573 \\
\hline
-1.58	 & 	\nodata	 & 	\nodata	 & 	\nodata	 & 	2.39	 & 	2.23	 & 	2.73	 & 	3.23	 & 	3.13	(-2.45) \\
-1.48	 & 	\nodata	 & 	\nodata	 & 	\nodata	 & 	2.40	 & 	2.30	 & 	2.80	 & 	3.30	 & 	3.19	(-2.45) \\
-1.37	 & 	\nodata	 & 	\nodata	 & 	\nodata	 & 	2.42	 & 	2.36	 & 	2.86	 & 	3.36	 & 	3.25	(-2.45) \\
-1.26	 & 	\nodata	 & 	\nodata	 & 	\nodata	 & 	2.44	 & 	2.44	 & 	2.94	 & 	3.44	 & 	3.31	(-2.45) \\
-1.15	 & 	\nodata	 & 	\nodata	 & 	\nodata	 & 	2.46	 & 	2.51	 & 	3.01	 & 	3.51	 & 	3.38	(-2.45) \\
-1.04	 & 	\nodata	 & 	\nodata	 & 	\nodata	 & 	2.48	 & 	2.60	 & 	3.10	 & 	3.60	 & 	3.46	(-2.45) \\
-0.93	 & 	\nodata	 & 	\nodata	 & 	\nodata	 & 	2.51	 & 	2.70	 & 	3.20	 & 	3.70	 & 	3.55	(-2.45) \\
-0.82	 & 	\nodata	 & 	\nodata	 & 	\nodata	 & 	2.53	 & 	2.81	 & 	3.31	 & 	3.81	 & 	3.65	(-2.45) \\
-0.71	 & 	\nodata	 & 	\nodata	 & 	\nodata	 & 	2.60	 & 	2.93	 & 	3.43	 & 	3.93	 & 	3.88	(-2.45) \\
-0.60	 & 	\nodata	 & 	\nodata	 & 	\nodata	 & 	2.64	 & 	3.08	 & 	3.58	 & 	4.08	 & 	4.02	(-2.45) \\
-0.50	 & 	2.44	 & 	3.14	 & 	2.20	 & 	2.68	 & 	3.25	 & 	3.75	 & 	4.25	 & 	4.19	(-2.45) \\
-0.38	 & 	2.56	 & 	3.66	 & 	2.80	 & 	2.75	 & 	3.47	 & 	3.97	 & 	4.47	 & 	4.41	(-2.45) \\
-0.28	 & 	\nodata	 & 	4.07	 & 	3.60	 & 	2.83	 & 	3.76	 & 	4.26	 & 	4.76	 & 	4.72	(-2.45) \\
-0.16	 & 	3.42	 & 	4.42	 & 	3.80	 & 	2.95	 & 	4.20	 & 	4.70	 & 	5.20	 & 	5.17	(-2.45) \\
-0.06	 & 	4.77	 & 	5.27	 & 	3.80	 & 	3.21	 & 	5.16	 & 	5.66	 & 	6.16	 & 	6.12	(-2.45) \\
0.06	 & 	4.37	 & 	5.07	 & 	3.90	 & 	3.03	 & 	5.16	 & 	5.66	 & 	6.16	 & 	6.12	(-2.45) \\
0.16	 & 	3.32	 & 	4.12	 & 	3.70	 & 	2.84	 & 	4.20	 & 	4.70	 & 	5.20	 & 	5.17	(-2.45) \\
0.28	 & 	3.37	 & 	3.77	 & 	2.60	 & 	2.75	 & 	3.76	 & 	4.26	 & 	4.76	 & 	4.72	(-2.45) \\
0.38	 & 	3.16	 & 	3.46	 & 	2.40	 & 	2.69	 & 	3.47	 & 	3.97	 & 	4.47	 & 	4.41	(-2.45) \\
0.50	 & 	2.44	 & 	3.14	 & 	2.20	 & 	2.65	 & 	3.25	 & 	3.75	 & 	4.25	 & 	4.19	(-2.45) \\
0.60	 & 	\nodata	 & 	\nodata	 & 	\nodata	 & 	2.62	 & 	3.08	 & 	3.58	 & 	4.08	 & 	4.02	(-2.45) \\
0.71	 & 	\nodata	 & 	\nodata	 & 	\nodata	 & 	2.59	 & 	2.93	 & 	3.43	 & 	3.93	 & 	3.88	(-2.45) \\
0.82	 & 	\nodata	 & 	\nodata	 & 	\nodata	 & 	2.54	 & 	2.81	 & 	3.31	 & 	3.81	 & 	3.65	(-2.45) \\
0.93	 & 	\nodata	 & 	\nodata	 & 	\nodata	 & 	2.52	 & 	2.70	 & 	3.20	 & 	3.70	 & 	3.55	(-2.45) \\
1.04	 & 	\nodata	 & 	\nodata	 & 	\nodata	 & 	2.50	 & 	2.60	 & 	3.10	 & 	3.60	 & 	3.46	(-2.45) \\
1.15	 & 	\nodata	 & 	\nodata	 & 	\nodata	 & 	2.49	 & 	2.51	 & 	3.01	 & 	3.51	 & 	3.38	(-2.45) \\
1.26	 & 	\nodata	 & 	\nodata	 & 	\nodata	 & 	2.47	 & 	2.44	 & 	2.94	 & 	3.44	 & 	3.31	(-2.45) \\
1.37	 & 	\nodata	 & 	\nodata	 & 	\nodata	 & 	2.46	 & 	2.36	 & 	2.86	 & 	3.36	 & 	3.25	(-2.45) \\
1.48	 & 	\nodata	 & 	\nodata	 & 	\nodata	 & 	2.45	 & 	2.30	 & 	2.80	 & 	3.30	 & 	3.19	(-2.45) \\
1.58	 & 	\nodata	 & 	\nodata	 & 	\nodata	 & 	2.44	 & 	2.23	 & 	2.73	 & 	3.23	 & 	3.13	(-2.45) \\
\hline
Mrk 78 \\
\hline
-3.00	 & 	0.93	 & 	1.93	 & 	3.13	 & 	\nodata	 & 	0.82	 & 	1.32	 & 	1.82	 & 	2.58	(-3.25) \\
-2.80	 & 	0.79	 & 	2.99	 & 	3.39	 & 	\nodata	 & 	0.88	 & 	1.38	 & 	1.88	 & 	2.59	(-3.20) \\
-2.73	 & 	\nodata	 & 	\nodata	 & 	\nodata	 & 	\nodata	 & 	0.91	 & 	1.41	 & 	1.91	 & 	\nodata	(\nodata) \\
-2.60	 & 	0.66	 & 	1.26	 & 	3.46	 & 	\nodata	 & 	0.95	 & 	1.45	 & 	1.95	 & 	2.61	(-3.15) \\
-2.40	 & 	0.93	 & 	1.33	 & 	3.53	 & 	\nodata	 & 	1.02	 & 	1.52	 & 	2.02	 & 	2.63	(-3.10) \\
-2.39	 & 	0.93	 & 	1.33	 & 	3.53	 & 	\nodata	 & 	1.02	 & 	1.52	 & 	2.02	 & 	2.63	(-3.10) \\
-2.20	 & 	\nodata	 & 	1.40	 & 	3.20	 & 	\nodata	 & 	1.09	 & 	1.59	 & 	2.09	 & 	2.65	(-3.05) \\
-2.04	 & 	\nodata	 & 	\nodata	 & 	\nodata	 & 	\nodata	 & 	1.16	 & 	1.66	 & 	2.16	 & 	\nodata	(\nodata) \\
-2.00	 & 	1.29	 & 	2.69	 & 	3.49	 & 	\nodata	 & 	1.18	 & 	1.68	 & 	2.18	 & 	2.64	(-2.95) \\
-1.80	 & 	0.58	 & 	1.78	 & 	2.78	 & 	3.11	 & 	1.27	 & 	1.77	 & 	2.27	 & 	2.63	(-2.85) \\
-1.70	 & 	\nodata	 & 	\nodata	 & 	\nodata	 & 	\nodata	 & 	1.31	 & 	1.81	 & 	2.31	 & 	\nodata	(\nodata) \\
-1.60	 & 	1.68	 & 	1.68	 & 	3.68	 & 	2.43	 & 	1.37	 & 	1.87	 & 	2.37	 & 	2.63	(-2.75) \\
-1.40	 & 	1.40	 & 	1.80	 & 	3.60	 & 	2.60	 & 	1.49	 & 	1.99	 & 	2.49	 & 	2.65	(-2.65) \\
-1.36	 & 	\nodata	 & 	\nodata	 & 	\nodata	 & 	\nodata	 & 	1.51	 & 	2.01	 & 	2.51	 & 	\nodata	(\nodata) \\
-1.20	 & 	\nodata	 & 	1.33	 & 	3.53	 & 	2.60	 & 	1.62	 & 	2.12	 & 	2.62	 & 	2.68	(-2.55) \\
-1.02	 & 	\nodata	 & 	\nodata	 & 	\nodata	 & 	\nodata	 & 	1.76	 & 	2.26	 & 	2.76	 & 	\nodata	(\nodata) \\
-1.00	 & 	1.49	 & 	2.09	 & 	4.09	 & 	2.39	 & 	1.78	 & 	2.28	 & 	2.78	 & 	2.74	(-2.45) \\
-0.80	 & 	1.88	 & 	2.28	 & 	3.68	 & 	2.60	 & 	1.97	 & 	2.47	 & 	2.97	 & 	2.83	(-2.35) \\
-0.68	 & 	\nodata	 & 	\nodata	 & 	\nodata	 & 	\nodata	 & 	2.11	 & 	2.61	 & 	3.11	 & 	\nodata	(\nodata) \\
-0.60	 & 	1.73	 & 	2.53	 & 	4.13	 & 	2.47	 & 	2.22	 & 	2.72	 & 	3.22	 & 	2.98	(-2.25) \\
-0.40	 & 	\nodata	 & 	2.88	 & 	4.08	 & 	3.14	 & 	2.57	 & 	3.07	 & 	3.57	 & 	3.23	(-2.15) \\
-0.34	 & 	\nodata	 & 	\nodata	 & 	\nodata	 & 	\nodata	 & 	2.71	 & 	3.21	 & 	3.71	 & 	\nodata	(\nodata) \\
-0.20	 & 	3.09	 & 	4.09	 & 	\nodata	 & 	3.06	 & 	3.18	 & 	3.68	 & 	4.18	 & 	3.74	(-2.05) \\
0.20	 & 	\nodata	 & 	4.09	 & 	5.89	 & 	2.73	 & 	3.18	 & 	3.68	 & 	4.18	 & 	3.74	(-2.05) \\
0.34	 & 	\nodata	 & 	\nodata	 & 	\nodata	 & 	\nodata	 & 	2.71	 & 	3.21	 & 	3.71	 & 	\nodata	(\nodata) \\
0.40	 & 	2.28	 & 	3.68	 & 	4.08	 & 	2.66	 & 	2.57	 & 	3.07	 & 	3.57	 & 	3.23	(-2.15) \\
0.60	 & 	2.13	 & 	2.53	 & 	3.73	 & 	2.61	 & 	2.22	 & 	2.72	 & 	3.22	 & 	2.98	(-2.25) \\
0.68	 & 	\nodata	 & 	\nodata	 & 	\nodata	 & 	\nodata	 & 	2.11	 & 	2.61	 & 	3.11	 & 	\nodata	(\nodata) \\
0.80	 & 	1.68	 & 	2.28	 & 	3.48	 & 	2.55	 & 	1.97	 & 	2.47	 & 	2.97	 & 	2.83	(-2.35) \\
1.00	 & 	2.09	 & 	2.09	 & 	3.29	 & 	2.63	 & 	1.78	 & 	2.28	 & 	2.78	 & 	2.74	(-2.45) \\
1.02	 & 	\nodata	 & 	\nodata	 & 	\nodata	 & 	\nodata	 & 	1.76	 & 	2.26	 & 	2.76	 & 	\nodata	(\nodata) \\
1.20	 & 	1.53	 & 	2.93	 & 	3.13	 & 	2.80	 & 	1.62	 & 	2.12	 & 	2.62	 & 	2.68	(-2.55) \\
1.36	 & 	\nodata	 & 	\nodata	 & 	\nodata	 & 	\nodata	 & 	1.51	 & 	2.01	 & 	2.51	 & 	\nodata	(\nodata) \\
1.40	 & 	\nodata	 & 	1.80	 & 	3.20	 & 	2.44	 & 	1.49	 & 	1.99	 & 	2.49	 & 	2.65	(-2.65) \\
1.60	 & 	\nodata	 & 	1.68	 & 	3.68	 & 	3.25	 & 	1.37	 & 	1.87	 & 	2.37	 & 	2.63	(-2.75) \\
1.70	 & 	\nodata	 & 	\nodata	 & 	\nodata	 & 	\nodata	 & 	1.31	 & 	1.81	 & 	2.31	 & 	\nodata	(\nodata) \\
1.80	 & 	\nodata	 & 	1.58	 & 	2.98	 & 	2.55	 & 	1.27	 & 	1.77	 & 	2.27	 & 	2.63	(-2.85) \\
2.00	 & 	0.89	 & 	1.49	 & 	3.29	 & 	2.60	 & 	1.18	 & 	1.68	 & 	2.18	 & 	2.64	(-2.95) \\
2.04	 & 	\nodata	 & 	\nodata	 & 	\nodata	 & 	\nodata	 & 	1.16	 & 	1.66	 & 	2.16	 & 	\nodata	(\nodata) \\
2.20	 & 	\nodata	 & 	1.40	 & 	3.20	 & 	2.95	 & 	1.09	 & 	1.59	 & 	2.09	 & 	2.65	(-3.05) \\
2.39	 & 	0.93	 & 	1.33	 & 	3.53	 & 	2.58	 & 	1.02	 & 	1.52	 & 	2.02	 & 	2.63	(-3.10) \\
2.40	 & 	0.93	 & 	1.33	 & 	3.53	 & 	2.58	 & 	1.02	 & 	1.52	 & 	2.02	 & 	2.63	(-3.10) \\
2.60	 & 	0.66	 & 	1.26	 & 	3.46	 & 	2.37	 & 	0.95	 & 	1.45	 & 	1.95	 & 	2.61	(-3.15) \\
2.73	 & 	\nodata	 & 	\nodata	 & 	\nodata	 & 	\nodata	 & 	0.91	 & 	1.41	 & 	1.91	 & 	\nodata	(\nodata) \\
2.80	 & 	0.79	 & 	2.99	 & 	3.39	 & 	2.54	 & 	0.88	 & 	1.38	 & 	1.88	 & 	2.59	(-3.20) \\
3.00	 & 	0.93	 & 	1.93	 & 	3.13	 & 	2.36	 & 	0.82	 & 	1.32	 & 	1.82	 & 	2.58	(-3.25) \\
\hline
Mrk 34 \\
\hline
-1.94	 & 	\nodata	 & 	1.46	 & 	3.46	 & 	2.19	 & 	1.25	 & 	1.75	 & 	2.25	 & 	\nodata	(\nodata) \\
-1.80	 & 	\nodata	 & 	\nodata	 & 	\nodata	 & 	\nodata	 & 	1.32	 & 	1.82	 & 	2.32	 & 	\nodata	(\nodata) \\
-1.68	 & 	\nodata	 & 	\nodata	 & 	\nodata	 & 	\nodata	 & 	1.38	 & 	1.88	 & 	2.38	 & 	\nodata	(\nodata) \\
-1.54	 & 	\nodata	 & 	1.64	 & 	3.34	 & 	2.45	 & 	1.46	 & 	1.96	 & 	2.46	 & 	\nodata	(\nodata) \\
-1.40	 & 	\nodata	 & 	\nodata	 & 	\nodata	 & 	\nodata	 & 	1.53	 & 	2.03	 & 	2.53	 & 	2.47	(-2.65) \\
-1.27	 & 	\nodata	 & 	\nodata	 & 	\nodata	 & 	2.99	 & 	1.62	 & 	2.12	 & 	2.62	 & 	2.53	(-2.65) \\
-1.14	 & 	\nodata	 & 	1.86	 & 	3.86	 & 	1.18	 & 	1.72	 & 	2.22	 & 	2.72	 & 	2.60	(-2.65) \\
-1.01	 & 	\nodata	 & 	\nodata	 & 	\nodata	 & 	\nodata	 & 	1.83	 & 	2.33	 & 	2.83	 & 	2.77	(-2.65) \\
-0.87	 & 	\nodata	 & 	\nodata	 & 	\nodata	 & 	2.25	 & 	1.95	 & 	2.45	 & 	2.95	 & 	2.87	(-2.65) \\
-0.74	 & 	\nodata	 & 	2.14	 & 	3.64	 & 	2.17	 & 	2.10	 & 	2.60	 & 	3.10	 & 	2.98	(-2.65) \\
-0.60	 & 	\nodata	 & 	\nodata	 & 	\nodata	 & 	\nodata	 & 	2.27	 & 	2.77	 & 	3.27	 & 	3.24	(-2.65) \\
-0.47	 & 	\nodata	 & 	\nodata	 & 	\nodata	 & 	\nodata	 & 	2.49	 & 	2.99	 & 	3.49	 & 	3.41	(-2.65) \\
-0.34	 & 	1.88	 & 	3.38	 & 	\nodata	 & 	2.51	 & 	2.78	 & 	3.28	 & 	3.78	 & 	3.63	(-2.65) \\
-0.20	 & 	\nodata	 & 	\nodata	 & 	\nodata	 & 	\nodata	 & 	3.22	 & 	3.72	 & 	4.22	 & 	3.94	(-2.65) \\
$\pm$0.14	 & 	1.59	 & 	3.49	 & 	\nodata	 & 	\nodata	 & 	3.58	 & 	4.08	 & 	4.58	 & 	4.39	(-2.65) \\
0.20	 & 	\nodata	 & 	\nodata	 & 	\nodata	 & 	\nodata	 & 	3.22	 & 	3.72	 & 	4.22	 & 	3.94	(-2.65) \\
0.34	 & 	1.28	 & 	3.28	 & 	\nodata	 & 	3.44	 & 	2.78	 & 	3.28	 & 	3.78	 & 	3.63	(-2.65) \\
0.47	 & 	\nodata	 & 	\nodata	 & 	\nodata	 & 	3.09	 & 	2.49	 & 	2.99	 & 	3.49	 & 	3.41	(-2.65) \\
0.60	 & 	\nodata	 & 	\nodata	 & 	\nodata	 & 	\nodata	 & 	2.27	 & 	2.77	 & 	3.27	 & 	3.24	(-2.65) \\
0.74	 & 	\nodata	 & 	2.14	 & 	3.34	 & 	3.13	 & 	2.10	 & 	2.60	 & 	3.10	 & 	2.98	(-2.65) \\
0.87	 & 	\nodata	 & 	\nodata	 & 	\nodata	 & 	2.89	 & 	1.95	 & 	2.45	 & 	2.95	 & 	2.87	(-2.65) \\
1.01	 & 	\nodata	 & 	\nodata	 & 	\nodata	 & 	\nodata	 & 	1.83	 & 	2.33	 & 	2.83	 & 	2.77	(-2.65) \\
1.14	 & 	\nodata	 & 	1.86	 & 	3.76	 & 	3.17	 & 	1.72	 & 	2.22	 & 	2.72	 & 	2.60	(-2.65) \\
1.27	 & 	\nodata	 & 	\nodata	 & 	\nodata	 & 	3.73	 & 	1.62	 & 	2.12	 & 	2.62	 & 	2.53	(-2.65) \\
1.40	 & 	\nodata	 & 	\nodata	 & 	\nodata	 & 	\nodata	 & 	1.53	 & 	2.03	 & 	2.53	 & 	2.47	(-2.65) \\
1.54	 & 	\nodata	 & 	1.64	 & 	3.44	 & 	2.55	 & 	1.46	 & 	1.96	 & 	2.46	 & 	\nodata	(\nodata) \\
1.68	 & 	\nodata	 & 	\nodata	 & 	\nodata	 & 	\nodata	 & 	1.38	 & 	1.88	 & 	2.38	 & 	\nodata	(\nodata) \\
1.80	 & 	\nodata	 & 	\nodata	 & 	\nodata	 & 	\nodata	 & 	1.32	 & 	1.82	 & 	2.32	 & 	\nodata	(\nodata) \\
1.94	 & 	\nodata	 & 	1.46	 & 	3.46	 & 	3.37	 & 	1.25	 & 	1.75	 & 	2.25	 & 	\nodata	(\nodata) \\
\enddata
\tablecomments{ A summary of the data presented in Figure~\ref{densities}. The first column lists the spatial distance from the nucleus in arcseconds, while Columns~2$-$4 provide the Cloudy model densities from the references listed in the Figure~\ref{densities} caption. The power-law density profiles (Columns~6$-$9) are evaluated on an evenly spaced grid, and extend beyond the radii with Cloudy models, thus there is not a matching Cloudy model density at every location. The radii correspond exactly to the values presented in Columns~6$-$9, while a Cloudy model component is shown if one exists within $<$0\farcs1 of the evaluated radius. The last column lists the variable power-law densities (see S\ref{ssec:uphotvar}), with the corresponding ionization parameters provided in parentheses.}
\label{densitytable}
\end{deluxetable*}
\textcolor{white}{.}
\clearpage

\subsection{[S~II] Ratio to Electron Density}\label{appendix2b}

This table provides the discrete values displayed in Figure~\ref{tempdengrid}, which are useful for converting [S~II] ratios to log($n_e$) values.

\begin{deluxetable}{cccc}[ht!]
\setlength{\tabcolsep}{0.11in}
\def\arraystretch{0.9}
\tablecaption{[S~II] Ratio to log($n_e$) Conversion\vspace{-0.5em}}
\tablehead{
\colhead{log($n_e$)} & \multicolumn{3}{c}{Value of [S~II] Ratio}\\
\cline{2-4}
\colhead{cm$^{-3}$} & \colhead{$T_e=$~10,000} & \colhead{$T_e=$~15,000} & \colhead{$T_e=$~20,000}
}
\startdata
1.0	 & 	1.420	 & 	1.400	 & 	1.380	 \\ 
1.1	 & 	1.420	 & 	1.390	 & 	1.380	 \\ 
1.2	 & 	1.410	 & 	1.390	 & 	1.380	 \\ 
1.3	 & 	1.400	 & 	1.380	 & 	1.370	 \\ 
1.4	 & 	1.390	 & 	1.380	 & 	1.370	 \\ 
1.5	 & 	1.380	 & 	1.370	 & 	1.360	 \\ 
1.6	 & 	1.370	 & 	1.360	 & 	1.350	 \\ 
1.7	 & 	1.360	 & 	1.350	 & 	1.340	 \\ 
1.8	 & 	1.340	 & 	1.330	 & 	1.330	 \\ 
1.9	 & 	1.320	 & 	1.310	 & 	1.310	 \\ 
2.0	 & 	1.290	 & 	1.290	 & 	1.290	 \\ 
2.1	 & 	1.260	 & 	1.260	 & 	1.270	 \\ 
2.2	 & 	1.220	 & 	1.230	 & 	1.240	 \\ 
2.3	 & 	1.180	 & 	1.200	 & 	1.210	 \\ 
2.4	 & 	1.130	 & 	1.150	 & 	1.170	 \\ 
2.5	 & 	1.080	 & 	1.110	 & 	1.130	 \\ 
2.6	 & 	1.030	 & 	1.060	 & 	1.080	 \\ 
2.7	 & 	0.969	 & 	1.000	 & 	1.030	 \\ 
2.8	 & 	0.911	 & 	0.949	 & 	0.976	 \\ 
2.9	 & 	0.853	 & 	0.892	 & 	0.921	 \\ 
3.0	 & 	0.798	 & 	0.836	 & 	0.865	 \\ 
3.1	 & 	0.746	 & 	0.782	 & 	0.810	 \\ 
3.2	 & 	0.698	 & 	0.731	 & 	0.757	 \\ 
3.3	 & 	0.655	 & 	0.684	 & 	0.708	 \\ 
3.4	 & 	0.617	 & 	0.643	 & 	0.664	 \\ 
3.5	 & 	0.585	 & 	0.606	 & 	0.624	 \\ 
3.6	 & 	0.557	 & 	0.574	 & 	0.589	 \\ 
3.7	 & 	0.533	 & 	0.547	 & 	0.559	 \\ 
3.8	 & 	0.513	 & 	0.524	 & 	0.533	 \\ 
3.9	 & 	0.497	 & 	0.505	 & 	0.512	 \\ 
4.0	 & 	0.484	 & 	0.489	 & 	0.494	 \\ 
4.1	 & 	0.473	 & 	0.476	 & 	0.480	 \\ 
4.2	 & 	0.464	 & 	0.466	 & 	0.468	 \\ 
4.3	 & 	0.457	 & 	0.457	 & 	0.458	 \\ 
4.4	 & 	0.452	 & 	0.450	 & 	0.450	 \\ 
4.5	 & 	0.447	 & 	0.445	 & 	0.444	 \\ 
4.6	 & 	0.443	 & 	0.441	 & 	0.439	 \\ 
4.7	 & 	0.441	 & 	0.437	 & 	0.435	 \\ 
4.8	 & 	0.438	 & 	0.434	 & 	0.432	 \\ 
4.9	 & 	0.437	 & 	0.432	 & 	0.430	 \\ 
5.0	 & 	0.435	 & 	0.431	 & 	0.428
\enddata
\tablecomments{The data for Figure~\ref{tempdengrid} for discrete temperatures (K). These values apply to optically-thick gas with 50\% ISM dust levels.}
\end{deluxetable}

\subsection{[O~III]/H\texorpdfstring{$\beta$}{B} to Ionization Parameter}\label{appendix2c}

This table provides the discrete values displayed in Figure~\ref{3do3}, which are useful for converting [O~III]/H$\beta$ ratios to log($U$).

\begin{deluxetable}{cccccc}[ht!]
\setlength{\tabcolsep}{0.025in}
\def\arraystretch{1.0}
\tablecaption{[O~III]/H$\beta$ to log($U$) Conversion\vspace{-0.5em}}
\tablehead{
\colhead{log($U$)} & \multicolumn{5}{c}{Value of [O~III]/H$\beta$ Ratio}\\
\cline{2-6}
\colhead{Param} & \colhead{$n_\mathrm{H}=10^2$} & \colhead{$n_\mathrm{H}=$~10$^3$} & \colhead{$n_\mathrm{H}=$~10$^4$} & \colhead{$n_\mathrm{H}=$~10$^5$} & \colhead{$n_\mathrm{H}=$~10$^6$}
}
\startdata
$-$6.0	 & 	$<$0.01	 & 	$<$0.01	 & 	$<$0.01	 & 	$<$0.01	 & 	$<$0.01	 \\ 
$-$5.5	 & 	$<$0.01	 & 	$<$0.01	 & 	$<$0.01	 & 	$<$0.01	 & 	$<$0.01	 \\ 
$-$5.0	 & 	$<$0.01	 & 	$<$0.01	 & 	$<$0.01	 & 	$<$0.01	 & 	$<$0.01	 \\ 
$-$4.5	 & 	0.01	 & 	0.02	 & 	0.03	 & 	0.04	 & 	0.03	 \\ 
$-$4.0	 & 	0.16	 & 	0.22	 & 	0.31	 & 	0.42	 & 	0.25	 \\ 
$-$3.5	 & 	1.62	 & 	2.09	 & 	2.83	 & 	3.42	 & 	2.12	 \\ 
$-$3.0	 & 	5.97	 & 	7.57	 & 	9.40	 & 	10.01	 & 	6.50	 \\ 
$-$2.5	 & 	10.81	 & 	13.00	 & 	15.32	 & 	15.64	 & 	9.97	 \\ 
$-$2.0	 & 	15.11	 & 	17.13	 & 	19.14	 & 	18.92	 & 	11.44	 \\ 
$-$1.5	 & 	19.06	 & 	20.24	 & 	21.16	 & 	19.73	 & 	10.87	 \\ 
$-$1.0	 & 	17.78	 & 	17.98	 & 	17.51	 & 	14.86	 & 	7.34	 \\ 
$-$0.5	 & 	8.09	 & 	7.98	 & 	7.48	 & 	6.09	 & 	3.19	 \\ 
$+$0.0	 & 	2.16	 & 	2.17	 & 	2.06	 & 	1.84	 & 	1.22	 \\ 
$+$0.5	 & 	1.11	 & 	1.12	 & 	1.12	 & 	1.10	 & 	1.03	 \\ 
$+$1.0	 & 	1.39	 & 	1.43	 & 	1.46	 & 	1.64	 & 	2.23	 
\enddata
\tablecomments{The data for Figure~\ref{3do3} for discrete densities (cm$^{-3}$). These values apply to optically-thick gas with 50\% ISM dust levels.}
\end{deluxetable}


\end{document}